\documentclass[11pt,a4paper]{article}
\usepackage{jhepmod}

\usepackage{color}			
\usepackage{amsbsy}		
\usepackage{amsfonts}		
\usepackage{amsmath}		
\usepackage{amssymb}		
\usepackage{color}
\usepackage{wasysym}
\usepackage{multirow}
\usepackage{psfrag}
\usepackage{slashed}
\usepackage[normalem]{ulem}

\newcommand*{\Ave}[1]{\mathinner{\left\langle{#1}\right\rangle}}

\newcommand{\GeV}{\;\mathrm{GeV}}
\newcommand{\TeV}{\;\mathrm{TeV}}

\newcommand{\FA}{{\tt FeynArts}}
\newcommand{\FC}{{\tt FormCalc}}

\newcommand{\gl}{\tilde{g}}
\newcommand{\sq}{\tilde{q}}
\newcommand{\st}{\tilde{t}}
\newcommand{\stb}{\tilde{\bar{t}}}
\newcommand{\sqb}{\tilde{\bar{q}}}
\newcommand{\neut}[1]{\tilde{\chi}^0_{#1}}
\newcommand{\cha}[1]{\tilde{\chi}^\pm_{#1}}
\newcommand{\SB}{{\mathrm{s}_\beta}}
\newcommand{\CB}{{\mathrm{c}_\beta}}

\newcommand{\reffig}[1]{{figure~\ref{fig:#1}}}
\newcommand{\refeq}[1]{{eq.~(\ref{eq:#1})}}

\title{Light NMSSM Higgs bosons in SUSY cascade decays at the LHC}

\ReportNo{DESY~11-128}

\author{Oscar~St{\aa}l}
\author{and Georg Weiglein}
\affiliation{%
 Deutsches Elektronen-Synchrotron DESY\\
 Notkestra{\ss}e 85, D--22607 Hamburg, Germany}

\emailAdd{oscar.stal@desy.de}
\emailAdd{georg.weiglein@desy.de}

\abstract{
An interesting feature of the next-to-minimal supersymmetric
standard model (NMSSM) is that one or more Higgs bosons may be
comparably light ($M_{H_i}<M_Z$) without being in conflict with
current experimental bounds. Due to a large singlet component, 
their direct production in standard channels at the Large Hadron
Collider (LHC) is suppressed. We demonstrate that there are good
prospects for observing such a light Higgs boson in decays
of heavy neutralinos and charginos. We consider an 
example scenario with $20\GeV < M_{H_1} < M_Z$ and show that a large
fraction of the cascade decays of gluinos and squarks involves the
production of at least one Higgs boson. 
Performing a Monte Carlo analysis at the level of fast detector
simulation, it is demonstrated how the Higgs signal can be separated
from the main backgrounds, giving access to the Yukawa coupling of the
Higgs to bottom quarks. Analyzing the resulting $b\bar{b}$ mass spectrum 
could provide an opportunity for light Higgs boson discovery already with 
$5$ fb$^{-1}$ of LHC data at $7$~TeV.}

\keywords{Supersymmetry, Higgs bosons, NMSSM, LHC}

\begin{document}

\maketitle
\section{Introduction}
The precise nature of the Higgs mechanism thought to be responsible for electroweak symmetry breaking remains unknown. To discover and study the properties of one or more Higgs bosons is therefore a challenge --- and one of the major objectives --- for the experiments at the running Large Hadron Collider (LHC). Most of the Higgs boson search strategies to date are designed to probe the Higgs sector of the Standard Model (SM), or its minimal supersymmetric extension (MSSM) \cite{Nilles:1983ge,Haber:1984rc}.  

There are several theoretically appealing arguments for weak-scale
supersymmetry to be realized in nature: it solves the hierarchy problem
of the SM Higgs mass, it enables gauge coupling unification, and with
R-parity conservation it also provides a natural dark matter candidate.
On the other hand, the realization of weak-scale supersymmetry in terms of the MSSM is not free of theoretical problems, such as the scale for the bilinear $\mu$-parameter entering the MSSM superpotential with positive mass dimension. This parameter has no natural values besides $M_{\mathrm{GUT}}$ or zero, while at the same time it must be close to the electroweak scale for a phenomenologically acceptable theory. To solve this problem in an elegant way the MSSM can be extended by a complex scalar singlet, giving the so-called next-to-minimal supersymmetric model (NMSSM). In this model an effective $\mu$-term of the right size can be generated dynamically from supersymmetry-breaking operators. For a general introduction to the NMSSM we refer to the recent reviews \cite{Ellwanger:2009dp,Maniatis:2009re}.

The NMSSM is characterized by an enlarged Higgs and neutralino sector as
compared to the MSSM, giving rise in particular to a richer Higgs
phenomenology. While it is well known that in certain scenarios of the
MSSM with complex parameters a light Higgs, with mass much below that of the
$Z$~boson, is unexcluded by the searches at LEP~\cite{Schael:2006cr} and
the Tevatron~\cite{Nakamura:2010zzi}
 (see 
\cite{Williams:2007dc,Williams:2011bu} for recent reevaluations with 
improved theoretical predictions), such a scenario can occur even more
generically in the NMSSM. In order to be compatible with the limits from
the LEP Higgs searches, in particular
the couplings of such a light Higgs state to
gauge bosons must be heavily suppressed. As a consequence of the
presence of a Higgs singlet, in the NMSSM such a situation happens
whenever a light Higgs state has a sufficiently large singlet component. 

The search for a heavier Higgs state with SM-like (or only moderately
suppressed) couplings to gauge bosons is complicated in such a scenario 
by the fact that often the decay of this heavier Higgs state 
into a pair of lighter Higgses is kinematically open, giving rise to
unusual decay signatures and to a large suppression of the standard
search channels for a SM-like Higgs. It should be noted in this context
that the observation of a decay of a heavier Higgs into a pair of
lighter Higgses would provide an opportunity for gaining experimental
access to triple-Higgs couplings, which are a crucial ingredient of 
electroweak symmetry breaking via the Higgs mechanism.

While in the NMSSM the case of a very light pseudo-scalar, 
$M_{A_ 1}<2m_b$, has found considerable attention in the literature, in
particular in the context of ``ideal'' Higgs scenarios \cite{Dermisek:2005ar,Dermisek:2005gg,Dermisek:2006wr,Dermisek:2007yt,Dermisek:2008uu,Dermisek:2010mg}, we will focus in
the following on scenarios with a light CP-even boson with 
$20\GeV < M_{H_1} < M_Z$. 
Within the MSSM the best known example of a light 
Higgs that is unexcluded by the present search limits is the 
``hole'' in the coverage of the CPX benchmark
scenario~\cite{Carena:2000ks} for 
$M_{H_1} \approx 45 \GeV$ and moderate values of 
$\tan\beta$~\cite{Schael:2006cr}, see \cite{Williams:2011bu} for a detailed 
discussion of the dependence of the unexcluded parameter region on the
choice of the various MSSM parameters. It will be difficult to cover
this parameter region with the standard
search channels at the 
LHC~\cite{Buescher:2005re,Schumacher:2004da,Accomando:2006ga}.
Various other (non-standard) search channels have been proposed which
may provide additional sensitivity in the quest to close
this ``CPX hole'' \cite{Akeroyd:2003jp,Ghosh:2004cc,Cheung:2007sva,Buescher:2005re,Carena:2007jk,Fowler:2009ay,Draper:2009au,Bandyopadhyay:2010tv,Bandyopadhyay:2011vc}.

In our analysis within the NMSSM we will investigate the prospects for
the production of light Higgs bosons in cascade decays of heavy SUSY
particles at the LHC. Such an analysis, where Higgs bosons are produced
in association with --- or in decays of --- other states of new physics,
is necessarily more model-dependent than the Higgs search in SM-like
channels. On the other hand, investigating Higgs physics in conjunction
with the production of other states of new physics offers additional
experimental opportunities and may also be more realistic, in the sense
that in order to extract a Higgs signal backgrounds both from SM-type and new
physics processes have to be considered. In the case of the MSSM with
real parameters, for a Higgs with a mass above the LEP limit for a SM
Higgs of $114.4\GeV$~\cite{Barate:2003sz}, a detailed experimental
study for Higgs boson production in a SUSY cascade has been carried out 
by the CMS Collaboration~\cite{Ball:2007zza}, involving a full detector
simulation and event reconstruction. 
These results, obtained for the benchmark point LM5,
cannot be directly translated to the case of searches for a Higgs boson with
mass far below $M_Z$, since in the latter case the $b$ jets resulting from 
the Higgs decay tend to be softer.
Further phenomenological analyses of Higgs production in
SUSY cascades in the MSSM with real parameters (and Higgs masses above
the LEP limits) have been carried out in 
\cite{Datta:2003iz,Huitu:2008sa,Gori:2011hj}, with recent developments focusing on jet substructure techniques to identify highly boosted Higgs bosons and enhance the discovery significance \cite{Kribs:2009yh,Kribs:2010hp}. 
The case of a lower mass Higgs has been considered in
\cite{Fowler:2009ay}, and it has been pointed out that in the CPX
scenario there is a significant rate for producing a light MSSM Higgs boson
in SUSY cascades, but no simulation of signal and background events was performed. The potential importance of SUSY cascades to establish a signal for a light CP-odd Higgs in the NMSSM has been pointed out in \cite{Cheung:2008rh}.

We generalize and extend the investigations carried out in
\cite{Fowler:2009ay,Cheung:2008rh} by calculating the sparticle decay modes in a
general NMSSM setting and performing a Monte
Carlo simulation of the signal and the dominant background to the level
of fast detector simulation. A simple cut-based analysis is performed to
demonstrate that signal and background can be resolved in the 
$b\bar{b}$+jets channel. The observation of the Higgs decay in 
the $b\bar{b}$ final state would be of interest also as a direct
manifestation of the Higgs Yukawa coupling.

The outline of our paper is as follows: the next section begins with a brief recapitulation of the NMSSM, presenting the scenario with a light CP-even  Higgs boson in some detail. In section 3, we describe the production of squarks and gluinos at the LHC and their eventual decay into Higgs bosons through electroweak cascades involving neutralinos and charginos. Section 4 describes a phenomenological Monte Carlo analysis of these cascade processes and contains the main results of this work in terms of kinematic distributions demonstrating the separation of signal and background. The conclusions are presented in section 5.

\section{The Next-to-Minimal Supersymmetric Standard Model}
In this section we review briefly the elements of the NMSSM which differ from the MSSM. Our conventions for the other sectors --- that remain unchanged when going to the NMSSM --- follow those of \cite{Skands:2003cj, SLHA2}.
\subsection{The Higgs sector}
\label{sect:higgs}
The $Z_3$-symmetric version of the NMSSM is given by the scale-invariant superpotential
\begin{equation}
W^{\mathrm{NMSSM}}=Y_u \hat{Q}_L\cdot{H_u}\hat{U}^c_L+Y_d \hat{Q}_L\cdot{H_d}\hat{D}^c_L+Y_e \hat{L}_L\cdot{H_d}\hat{E}^c_L+\lambda \hat{S} \hat{H}_u\cdot \hat{H}_d + \frac{1}{3}\kappa\hat{S}^3,
\end{equation}
where $\hat{\Phi}$ denotes a chiral superfield with scalar component $\Phi$. The complex gauge singlet $\hat{S}$ is a new addition with respect to the MSSM. To have a complete phenomenological model the soft SUSY-breaking terms must also be specified. These are extended by couplings of the singlet field, giving new contributions to the scalar potential
\begin{equation}
V^{\mathrm{NMSSM}}=V^{\mathrm{MSSM}}+m_S^2 S^2+\lambda A_\lambda S H_u\cdot H_d+\frac{1}{3}\kappa A_\kappa S^3.
\label{eq:Higgspot}
\end{equation}
The NMSSM Higgs potential, which is derived from the usual $F$-terms, $D$-terms and the soft-breaking potential given by Equation \eqref{eq:Higgspot}, allows for a minimum where the singlet develops a vacuum expectation value (vev) $v_s=\Ave{S}$. This induces an effective bilinear term $\lambda\Ave{S}H_u\cdot H_d$, thus providing a dynamical explanation for the $\mu$ parameter of the MSSM in terms of $\mu_\mathrm{eff}=\lambda v_s$.

Electroweak symmetry breaking (EWSB) proceeds similarly to the MSSM, and the two Higgs doublets are expanded around the potential minimum according to
\begin{equation}
H_d = \left(\begin{array}{c}
v_d+\frac{1}{\sqrt{2}}\left(\phi_d-\mathrm{i}\sigma_d\right)\\
-\phi_d^-
\end{array}\right),
\qquad
H_u = \left(\begin{array}{c}
\phi_u^+\\
v_u+\frac{1}{\sqrt{2}}\left(\phi_u+\mathrm{i}\sigma_u\right)
\end{array}\right).
\end{equation}
Equivalently, the singlet field has an expansion
\begin{equation}
S=v_s+\frac{1}{\sqrt{2}}\left(\phi_s+\mathrm{i}\sigma_s\right).
\end{equation} 
Using the minimization conditions for the potential, the scalar mass parameters $m_{H_u}^2$, $m_{H_d}^2$, and $m_{S}^2$ can be traded for
\begin{align*}
M_Z^2&=g^2 v^2=\frac{1}{2}\left(g_1^2+g_2^2\right)v^2 , \\
\tan\beta&=v_u/v_d , \\
\mu_{\mathrm{eff}}&=\lambda v_s,
\end{align*}
where the doublet vevs fulfill $v^2 \equiv v_u^2+v_d^2=(174\GeV)^2$.
Assuming a CP-invariant Higgs sector, all parameters are taken to be
real. The number of parameters is increased from the MSSM. In addition
to $M_A$ (or $M_{H^\pm}$), and $\tan\beta$, the values of $\lambda$,
$\kappa$, and $A_\kappa$ can be chosen as free parameters.

After EWSB, the addition of a complex scalar field gives rise to
additional particles in the NMSSM spectrum with respect to the MSSM: two
additional Higgs bosons (one of which is CP-even, the other CP-odd) and their fermionic partner, the singlino. The elements of the tree-level mass matrix $\mathcal{M}_H^2$ for the CP-even Higgs bosons are given in the basis $\left(\phi_d,\phi_u,\phi_s\right)$ by
\begin{equation}
\begin{aligned}
\left(\mathcal{M}_H^2\right)_{11} &=
M_Z^2\cos^2\beta+B\mu_{\mathrm{eff}}\tan\beta , \\
\left(\mathcal{M}_H^2\right)_{22} &=
M_Z^2\sin^2\beta+B\mu_{\mathrm{eff}}\cot\beta , \\
\left(\mathcal{M}_H^2\right)_{33} &= \frac{\lambda v^2 A_\lambda}{v_s}
\cos\beta\sin\beta+\kappa v_s\left(A_\kappa+4\kappa v_s\right) , \\
\left(\mathcal{M}_H^2\right)_{12} &= \left(2\lambda^2
v^2-M_Z^2\right)\cos\beta\sin\beta-B\mu_{\mathrm{eff}} , \\
\left(\mathcal{M}_H^2\right)_{13} &= \lambda
v\left[2\mu_{\mathrm{eff}}\cos\beta-\left(B+\kappa
v_s\right)\sin\beta\right] , \\
\left(\mathcal{M}_H^2\right)_{23} &= \lambda
v\left[2\mu_{\mathrm{eff}}\sin\beta-\left(B+\kappa
v_s\right)\cos\beta\right] ,
\end{aligned}
\end{equation}
where $B\equiv A_\lambda+\kappa v_s$. This matrix is diagonalized by a real $3\times 3$ matrix with elements $S_{ij}$, such that the Higgs mass eigenstates $H_i$ are given by $H_i=S_{ij}\phi_j$. For the CP-odd states, the mass matrix elements in the basis $\left(\sigma_d,\sigma_u,\sigma_s\right)$ can be written
\begin{equation}
\begin{aligned}
\left(\mathcal{M}_A^2\right)_{11} &=B\mu_{\mathrm{eff}}\tan\beta , \\
\left(\mathcal{M}_A^2\right)_{22} &=B\mu_{\mathrm{eff}}\cot\beta , \\
\left(\mathcal{M}_A^2\right)_{33} &=\frac{\lambda
v^2}{v_s}\left(B+3\kappa v_s\right)\cos\beta\sin\beta-3\kappa A_\kappa
v_s , \\
\left(\mathcal{M}_A^2\right)_{12} &=B\mu_{\mathrm{eff}} , \\
\left(\mathcal{M}_A^2\right)_{13} &=\lambda v\left(B-3\kappa
v_s\right)\sin\beta , \\
\left(\mathcal{M}_A^2\right)_{23} &=\lambda v\left(B-3\kappa
v_s\right)\cos\beta .
\end{aligned}
\end{equation}
Similarly to the CP-even case, the massive eigenstates $A_i$ can be
written using a mixing matrix $P_{ij}$ as $A_i=P_{ij}\sigma_j$. One
degree of freedom is massless and corresponds to the neutral Goldstone
boson providing the longitudinal degree of freedom to the $Z$ boson. For some purposes it can be convenient to introduce the CP-odd mass parameter 
\begin{equation}
M_A^2=\frac{B\mu_{\mathrm{eff}}}{\CB\SB},
\end{equation}
which corresponds to the mass of the CP-odd Higgs boson in the MSSM limit of the NMSSM.\footnote{The MSSM limit is obtained by taking $\lambda\to 0$, $\kappa\to 0$, while keeping the ratio $\kappa/\lambda$ and all dimensionful parameters fixed.} 

No additional charged scalar is introduced in the NMSSM, but the relation of the physical charged Higgs boson  mass to the CP-odd mass parameter gets modified. At tree-level the charged Higgs mass is now given by 
\begin{equation}
M_{H^\pm}^2=M_A^2+M_W^2-\lambda^2v^2.
\end{equation}

\subsection{The neutralino sector}

As a result of introducing the new singlet superfield $\hat{S}$, the NMSSM comes with an additional fermion partner of the complex scalar $S$, the singlino. The singlino mixes with the existing four neutralinos of the MSSM. 
The resulting $5\times 5$ mass matrix derives from the bilinear terms
\begin{equation}
\mathcal{L}=-\frac{1}{2}(\tilde{\psi}^0)^T {\cal M}_{\tilde{\chi}^0}\tilde{\psi}^0+\mathrm{h.c.},
\end{equation}
and in the basis $(-\mathrm{i}\tilde{B}, -\mathrm{i}\tilde{W}, \tilde{H}_d, \tilde{H}_u, 
\tilde{S})$ 
it is given by
\begin{equation}
\mathcal{M}_{\tilde{\chi}^0}=\left(
\begin{array}{ccccc}
\displaystyle M_1 & 0  & -\frac{g_1 v_d}{\sqrt{2}} & \frac{g_1 v_u}{\sqrt{2}} &  0\\
			    0 & M_2  & \frac{g_2 v_d}{\sqrt{2}}& -\frac{g_2 v_u}{\sqrt{2}} & 0\\
 -\frac{g_1 v_d}{\sqrt{2}} &\frac{g_2 v_d}{\sqrt{2}} & 0 & -\mu_{\mathrm{eff}} & -\lambda v_u \\
  \frac{g_1 v_u}{\sqrt{2}} & -\frac{g_2 v_u}{\sqrt{2}} &-\mu_{\mathrm{eff}} & 0 &-\lambda v_d \\
 0 & 0 & -\lambda v_u & -\lambda v_d & 2\kappa v_s
\end{array}
\right).
\label{eq:mneut}
\end{equation}
The upper left $4\times 4$ submatrix is identical to the neutralino mass matrix in the MSSM. The neutralino masses can be diagonalized by a single unitary matrix $N$ such that 
\begin{equation}
D=\mathrm{diag}(m_{\tilde{\chi}^0_i}) = N^* \mathcal{M}_{\tilde{\chi}^0} N^\dagger
\end{equation}
is real and positive with the neutralino mass eigenvalues in ascending order. Alternatively one can use a real mixing matrix $N$, and allow $D$ to have negative elements. In this case the physical neutralino masses are given by $|m_{\tilde{\chi}_i^0}|$ and the neutralino couplings incorporate the additional phase shift on the neutralino fields.

\subsection{The squark sector}
We adopt a universal value $M_{\mathrm{SUSY}}$ for the soft SUSY-breaking scalar mass parameters. This means that, for each squark pair $\tilde{q}_L$, $\tilde{q}_R$ of a given flavour, the mass matrix attains the form
\begin{equation}
\mathcal{M}^2_{\tilde{q}}=\left(
\begin{array}{cc}
\displaystyle M_{\mathrm{SUSY}}^2+m_q^2+M_Z^2\cos 2\beta(I_3^q-Q_q s_W^2) & m_q X_q \\
m_qX_q & M_{\mathrm{SUSY}}^2+m_q^2+M_Z^2\cos 2\beta Q_q s_W^2 \\
\end{array}
\right).
\label{eq:squarkmassmatrix}
\end{equation}
Here $m_q$ is the mass of the corresponding quark, $I_3^q$ the third
component of the weak isospin, and $Q_q$ the electric charge quantum
number. For the weak mixing angle we introduce the short-hand notations
$s_W\equiv \sin\theta_W$ and $c_W\equiv \cos\theta_W$. The off-diagonal
elements of $\mathcal{M}^2_{\tilde{q}}$ are related to the soft
trilinear couplings $A_q$ as $X_q=A_q-\mu_{\mathrm{eff}}\cot\beta$ for
up-type squarks, and $X_q=A_q-\mu_{\mathrm{eff}}\tan\beta$ for the case
of down-type squarks, respectively.  The mass eigenstates ($\tilde{q}_1$, $\tilde{q}_2$) are obtained by a diagonalization of the mass matrix. A generic squark mass will be denoted $M_{\tilde{q}}$ below.

\subsection{Scenarios with light Higgs bosons}

As mentioned above, we will focus in the following on the case where the
lightest CP-even Higgs boson of the NMSSM, $H_1$, has a mass much below
$M_Z$. The fact that such a light Higgs, possessing a heavily suppressed
coupling to gauge bosons as compared to the Higgs boson of the SM, may
be unexcluded by the current search limits is known already from the case
of the MSSM with complex 
parameters~\cite{Schael:2006cr,Williams:2007dc,Williams:2011bu}. 
In the NMSSM such a situation happens more generically, in particular
also for the case where the SUSY parameters are real. If the mass
eigenstate $H_1$ has a large component of the singlet interaction state
$\phi_s$, its couplings to gauge bosons (and also to quarks) will be
correspondingly suppressed. We will investigate the prospects for
detecting such a light Higgs state through its production in SUSY
cascades.

In the numerical analysis, we shall use a scenario derived from the
``P4'' benchmark point defined in \cite{Djouadi:2008uw}. This benchmark
can be realized in models with non-universal Higgs mass parameters
($m_{H_u}\neq m_{H_d}$) at the scale of grand unification, and it is
compatible with the data on the cold dark matter density. As originally
defined, the P4 benchmark contains a very light CP-even Higgs boson
($M_{H_1}=32.3\GeV$). In order to explore the full range $M_{H_1}< M_Z$, 
we slightly modify the scenario to allow changing the value of
$M_{H_1}$, with the remaining phenomenology essentially unchanged. To
this end we set $\lambda=0.6$ and allow $A_{\kappa}$ to take on values
in the range $0\GeV < A_\kappa < 300\GeV$.\footnote{The original P4
benchmark point is recovered for $\lambda=0.53$ and $A_\kappa =
220\GeV$.} The soft SUSY-breaking parameters are defined directly at the
SUSY-breaking scale, 
allowing us to consider a more general spectrum for the remaining
(non-Higgs) sectors of the theory. Values for the tree-level parameters
in the Higgs sector and the soft SUSY-breaking parameters in the
modified P4 scenario are specified in table~\ref{tab:param}. The two
values given for $M_{\mathrm{SUSY}}$ will be used later in the
phenomenological analysis, while the values quoted in this and 
the next section (unless otherwise stated) have been evaluated for 
$M_{\mathrm{SUSY}}=750\GeV$.
\begin{table}
\centering
\begin{tabular}{p{0.8cm}rp{1.5cm}p{0.8cm}rl}
\hline

\multicolumn{6}{c}{Higgs sector parameters}\\
$\lambda$ & $0.6$ & & $\kappa$ & $0.12$ \\
$\tan\beta$ & 2.6 & & $\mu_{\mathrm{eff}}$ & $-200$ & GeV  \\
$A_\lambda$ & $-510$ & GeV & $A_\kappa$ & $0$ -- $300$ & GeV \\
\hline
\multicolumn{6}{c}{Gaugino masses}\\
$M_1$ & $300$ & GeV & $M_2$ & $600$ & GeV \\
$M_3$ & $1000$ & GeV \\
\hline
\multicolumn{6}{c}{Trilinear couplings}\\
\multicolumn{6}{c}{$A_t=A_b=A_\tau=0 \GeV$} \\
\hline
\multicolumn{6}{c}{Soft scalar mass}\\
\multicolumn{6}{c}{$M_{\mathrm{SUSY}}=750 \GeV$, $1\TeV$} \\
\hline
\end{tabular}
\caption{Values for the NMSSM input parameters at the SUSY-breaking scale in the modified P4 scenario.}
\label{tab:param}
\end{table}

\begin{figure}
\centering
\includegraphics[width=70mm]{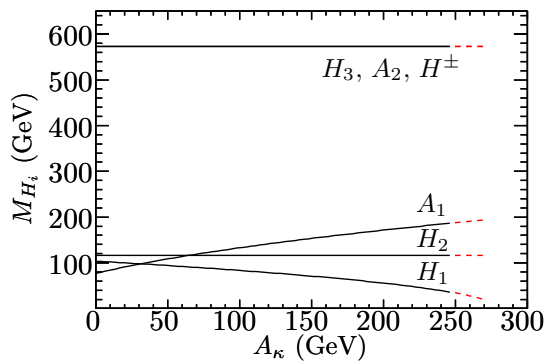}
\caption{Higgs mass spectrum for the modified P4 scenario with $M_{\mathrm{SUSY}}=750\GeV$ as a function of the free parameter $A_\kappa$. For $A_\kappa\gtrsim 250\GeV$ (dashed) the electroweak symmetry remains unbroken in the global minimum.}
\label{fig:hmass_Akappa}
\end{figure}

The NMSSM Higgs masses are subject to sizable corrections beyond leading
order \cite{Ellwanger:1993hn, Elliott:1993ex,
Elliott:1993uc,Elliott:1993bs,Pandita:1993hx,Ellwanger:2005fh}. In order 
to incorporate the most accurate predictions currently available \cite{Degrassi:2009yq}, {\tt NMSSMTools 2.3.5} \cite{Ellwanger:2004xm,Ellwanger:2005dv,Belanger:2005kh} is used to compute the Higgs spectrum. The resulting Higgs masses in the modified P4 scenario are shown in \reffig{hmass_Akappa} as a function of the free parameter $A_\kappa$.
In the region with $A_\kappa\gtrsim 250\GeV$ the global minimum of the
Higgs potential does not break the electroweak symmetry; hence these
values will not be considered. The masses of two Higgs bosons show a
dependence on $A_\kappa$: the lightest CP-even Higgs $H_1$ ($M_{H_1}$
varying from about $110\GeV$ to $20\GeV$), and the lightest CP-odd Higgs
$A_1$ (with $M_{A_1}$ going from about $90\GeV$ to $200\GeV$). $H_2$ is
always SM-like and has a mass of about $M_{H_2}=115\GeV$ ($118\GeV$ for $M_{\mathrm{SUSY}}=1\TeV$). For all Higgs masses in this plot the NMSSM scenario is compatible with the direct limits from Higgs searches. The light CP-even Higgs 
($M_{H_1}\ll M_Z$) is allowed due to a large singlet component, with
$|S_{13}|^2$ ranging from $0.9$ for $A_\kappa=0\GeV$ to
$|S_{13}|^2>0.99$ for $A_\kappa=250\GeV$. As a consequence, the couplings of $H_1$ to vector bosons are heavily suppressed, so that the cross section for production through Higgsstrahlung drops below the LEP limit. 
The pair production of $A_1H_1$ is even further suppressed by the large singlet fractions of both $H_1$ and $A_1$, while production of $H_2 A_1$ and $H_2 Z$ are beyond the kinematic reach of LEP.
 The full mass ranges shown in the figure are also compatible with the constraints from $B$-physics implemented in {\tt NMSSMTools 2.3.5} \cite{Ellwanger:2004xm,Ellwanger:2005dv,Belanger:2005kh}, as expected when the charged Higgs boson is heavy \cite{Mahmoudi:2010xp,Eriksson:2008cx}. The precise values obtained here for the heavy Higgs masses are $M_{H^\pm}\simeq 563\GeV$, and {$M_{H_3}\simeq M_{A_2}\simeq 572\GeV$. 
None of the heavy Higgs bosons will play any role in the following. With
the negative sign for the effective $\mu$ parameter, this model cannot
be used to explain the observed deviation in the anomalous magnetic
moment of the muon (see e.g.\ \cite{Stockinger:2006zn}
for a review). However --- since the considered value of $\tan\beta$ is 
rather low --- the predicted value for $(g-2)_{\mu}$ at least stays close to 
that in the SM. 
 
The branching ratios of the three lightest Higgs states, $H_1$, $H_2$, and $A_1$, are given in~\reffig{hdecay}.
\begin{figure}
\centering
\includegraphics[width=0.48\columnwidth]{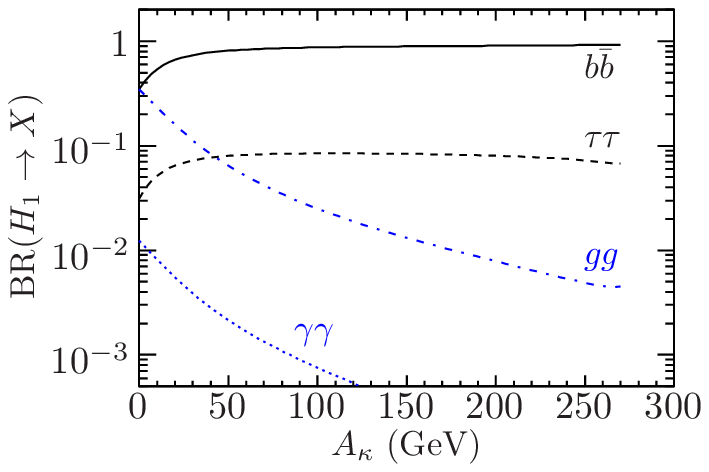}
\includegraphics[width=0.48\columnwidth]{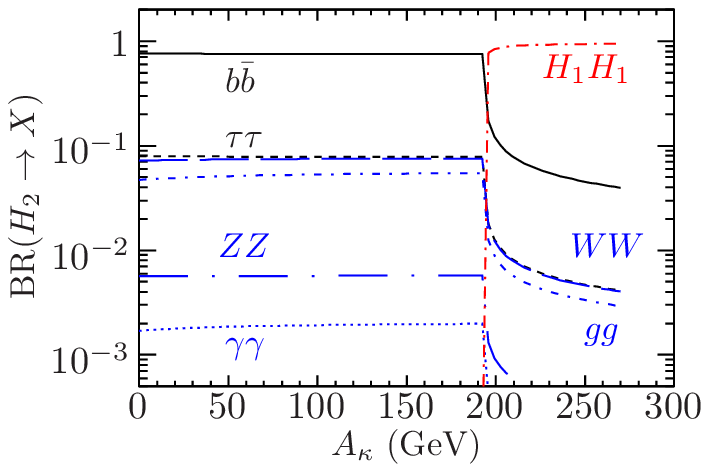}\\
\vspace{5mm}
\includegraphics[width=0.48\columnwidth]{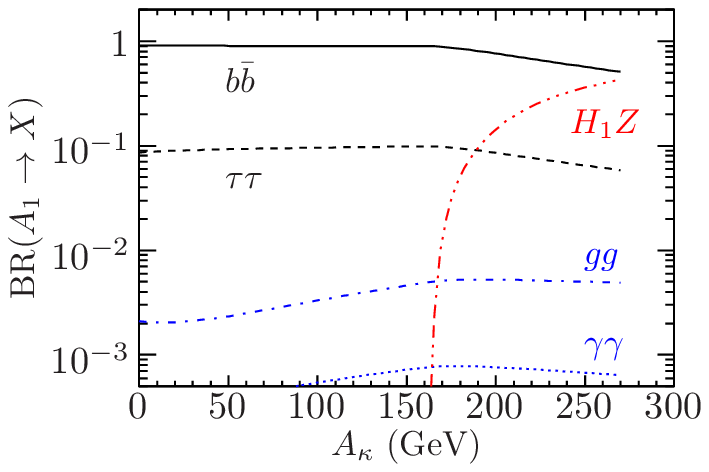}
\caption{Branching ratios of $H_1$ (upper left), $H_2$ (upper right), and $A_1$ (lower) in the modified P4 scenario with $M_{\mathrm{SUSY}}=750\GeV$.}
\label{fig:hdecay}
\end{figure}
As can be seen from this figure, the light singlet $H_1$ decays preferentially into $b\bar{b}$, with $\mathrm{BR}(H_1\to b\bar{b})\simeq 90\%$ over the full mass range. The subdominant decay into $\tau\tau$ basically saturates the $H_1$ width. For lower values of $A_\kappa$ --- where $M_{H_1}\gtrsim 90\GeV$ --- we note a similar enhancement of $\mathrm{BR}(H_1\to \gamma\gamma)$ compared to a SM Higgs with the same mass as recently discussed in \cite{Ellwanger:2010nf}. 
The $H_2$ has a more complicated decay pattern, in particular for low
$A_\kappa$ where $H_2\to b\bar{b}$ dominates and several competing modes
($H_2\to \tau\tau$, $gg$, $WW$) each have a branching fraction around
$10\%$. In this region $H_2$ is SM-like, and the same search strategies
as devised for the SM Higgs (and the lightest MSSM Higgs boson in the
decoupling limit) should apply. This situation changes radically when the channel $H_2\to H_1 H_1$ opens. When this is the case, the $H_2\to H_1 H_1$ mode becomes completely dominant. Finally, the lightest CP-odd Higgs $A_1$ decays predominantly into $b\bar{b}$, with a large fraction going into the mode $A_1\to H_1Z$ when kinematically accessible.

In the neutralino sector the mass spectrum is independent of $A_\kappa$
(and $M_{\mathrm{SUSY}}$) at lowest order, cf.~equation~\eqref{eq:mneut}, and therefore remains fixed at:
\begin{equation*}
\begin{aligned}
M_{\neut{1}}=97.6\GeV,&\quad M_{\neut{2}}=227\GeV,\quad M_{\neut{3}}=228\GeV \\
M_{\neut{4}}=304\GeV,&\quad M_{\neut{5}}=616\GeV.
\end{aligned}
\end{equation*}
There is a clear hierarchy in the mass parameters, which leads to a
small mixing between the neutralinos. The heaviest neutralino is almost
exclusively wino, and $\neut{4}$ is mostly bino. The intermediate mass states $\neut{2}$, and $\neut{3}$ are predominantly Higgsino, while the lightest neutralino $\neut{1}$ is the singlino. The lightest neutralino is also the overall lightest supersymmetric particle (LSP) in these scenarios and thereby a candidate for cold dark matter.

\section{Higgs production in the light $H_1$ scenario}
\label{sect:Higgsprod}
\subsection{Standard channels}
The rate for direct production of a light singlet $H_1$ in gluon fusion, $gg\to H_1$, is proportional to its reduced (squared) coupling to quarks.
Compared to a SM Higgs boson with the same mass, the dominant top loop
contribution contains the additional factor $|S_{12}|^2/\sin^2\beta$.
The size of $|S_{12}|^2$ is limited from above by 
$|S_{12}|^2\leq 1-|S_{13}|^2$, where $S_{13}$ is the singlet component.
For $M_{H_1}\ll M_Z$, where $|S_{13}|\to 1$, the rate for this process
gets heavily suppressed. The cross section for $H_1$ in weak boson
fusion, involving the coupling of $H_1$ to gauge bosons, is similarly 
suppressed.
In scenarios where $M_{H_1}> M_Z$ (corresponding to the mass range
below the LEP limit on a SM-like Higgs which is unexcluded in the MSSM
with real parameters) the suppression of $gg\to H_1$ can be overcome 
by an increased branching ratio for $H_1\to \gamma\gamma$~\cite{Ellwanger:2010nf}.

For $A_\kappa\gtrsim 200\GeV$ in the modified P4 scenario, $H_1$ is
light enough to be produced through the decay of the SM-like $H_2\to
H_1H_1$, which can be dominant, see~figure \ref{fig:hdecay}. The production of $H_2$ in standard channels is not suppressed.
The resulting two-step decay chain leads to ``unusual'' final states for 
$H_2$: $4b$ (about $82\%$ of all decays), $2b2\tau$ ($17\%$), 
and $4\tau$ ($0.6\%$). These final states make it difficult to establish
a Higgs signal, as it has been demonstrated, for instance, by the numerous 
attempts \cite{Ellwanger:2001iw,Ellwanger:2003jt,Ellwanger:2004gz,Forshaw:2007ra,Belyaev:2008gj} to establish a ``no-loose'' theorem for NMSSM Higgs 
searches when decays of the SM-like Higgs into lighter Higgses are open. 

Another possibility to produce $H_1$ in Higgs decays would be through 
the decay $A_1\to H_1 Z$. However, the singlet nature of $A_1$ in the 
modified P4 scenario leads to a suppression of $A_1$ production similar 
to that for $H_1$, and this mode is therefore not likely to be accessible.

The direct production of the heavy Higgs bosons $H_3$, $A_2$, and
$H^\pm$ is in principle not suppressed with respect to the MSSM case, but 
at a mass close to $600\GeV$ and low $\tan\beta$ the observation of
those states at the LHC will be difficult even at high luminosity. A 
large fraction of the heavy Higgs bosons in this scenario will decay 
into lighter Higgs bosons, neutralinos and charginos. A detailed 
investigation of these channels could possibly be of interest for a 
study assuming a very high luminosity at 14 TeV, but is beyond the scope of 
the present paper.

In summary, it will be problematic to produce and reconstruct the light 
$H_1$ in any of the standard channels proposed for Higgs production at 
the LHC. We shall focus instead on the possibility to produce $H_1$ 
in the decays of supersymmetric particles.

\subsection{SUSY cascades}
As discussed in the previous section, inclusive production of the
heavier state $H_2$ with subsequent decay $H_2\to H_1H_1$ may be
difficult to observe at the LHC. However, the related process where 
a heavier neutralino decays into a lighter neutralino and a Higgs boson 
(and the corresponding mode of the decay of the heavier chargino) may 
offer better
prospects. In fact, a light Higgs boson in the mass range below $M_Z$
may occur in a large fraction of cascade decays of heavier SUSY particles
that are produced via strong interaction processes. The hard scale 
associated with the sparticle production can lead to event signatures 
which are more clearly separable from the SM backgrounds than those 
of inclusive Higgs production followed by a decay into a pair of 
$H_1$ states. The processes of interest are 
\begin{align}
\neut{i}\to \neut{j} H_k,&\label{eq:chi0dec}\quad \neut{i}\to \neut{j} A_k,\\
\cha{2}\to \cha{1}H_k\label{eq:chipdec},&\quad \cha{2}\to \cha{1}A_k,
\end{align}
where $H_k$ ($A_k$) denotes any of the CP-even (CP-odd) Higgs bosons.
As mentioned above we do not consider scenarios where the heavier $H^\pm$ is produced in the cascades. The partial width for the neutralino decay \eqref{eq:chi0dec} is given at tree-level by
\begin{equation}
\Gamma(\neut{i}\to \neut{j} H_k)=\frac{|\mathcal{S}_{ijk}|^2}{16\pi m_{\neut{i}}^3}\tau^{1/2}(m_{\neut{i}}^2,m_{\neut{j}}^2,m_{H_k}^2)\left(m_{\neut{i}}^2+m_{\neut{j}}^2-m_{H_k}^2+2m_{\neut{i}}m_{\neut{j}}\right),
\label{eq:gachih}
\end{equation}
with a CP-even Higgs in the final state and
\begin{equation}
\Gamma(\neut{i}\to \neut{j} A_k)=\frac{|\mathcal{P}_{ijk}|^2}{16\pi m_{\neut{i}}^3}\tau^{1/2}(m_{\neut{i}}^2,m_{\neut{j}}^2,m_{A_k}^2)\left(m_{\neut{i}}^2+m_{\neut{j}}^2-m_{A_k}^2-2m_{\neut{i}}m_{\neut{j}}\right)
\label{eq:gachia}
\end{equation}
for the decay into a CP-odd scalar. The K{\"a}ll{\'e}n function $\tau(x,y,z)=(x-y-z)^2-4yz$, and the coupling factors are
\begin{equation}
\begin{aligned}
\mathcal{S}_{ijk}=&\frac{e}{2c_W s_W}\Bigl[\left(S_{k1}N_{i3}-S_{k2}N_{i4}\right)\left(c_WN_{j2}-s_WN_{j1}\right)\Bigr]-\\
&\frac{\lambda}{\sqrt{2}}\Bigl[N_{i5}\left(S_{k1}N_{j4}+S_{k2}N_{j3}\right)+S_{k3}N_{i4}N_{j3}\Bigr]+\frac{\kappa}{\sqrt{2}}S_{k3}N_{i5}N_{j5}+i \leftrightarrow j,
\end{aligned}
\end{equation}
and
\begin{equation}
\begin{aligned}
-\mathrm{i}\mathcal{P}_{ijk}=&\frac{e}{2c_W s_W}\Bigl[\left(P_{k1}N_{i3}-P_{k2}N_{i4}\right)\left(c_WN_{j2}-s_WN_{j1}\right)\Bigr]+\\
&\frac{\lambda}{\sqrt{2}}\Bigl[N_{i5}\left(P_{k1}N_{j4}+P_{k2}N_{j3}\right)+P_{k3}N_{i4}N_{j3}\Bigr]-\frac{\kappa}{\sqrt{2}}P_{k3}N_{i5}N_{j5}+i \leftrightarrow j,
\end{aligned}
\label{eq:Pijk}
\end{equation}
where the mixing matrices $S_{ij}$, $P_{ij}$, and $N_{ij}$ are defined
in section \ref{sect:higgs}. Equations
\eqref{eq:gachih}--\eqref{eq:Pijk} assume a real neutralino mixing
matrix $N_{ij}$ and signed neutralino masses. Competing neutralino decay
modes are into vector bosons, $\neut{i}\to \neut{j} Z$ and $\neut{i}\to
\cha{j} W^\mp$. 
For brevity we refrain from giving expressions for these (and the corresponding chargino decay modes) here; they can be found in \cite{Choi:2004zx}. A detailed analysis of the $W^\pm$ mode is performed in \cite{Liebler:2010bi}. Since the squarks and sleptons are assumed to be heavy, there are no open two-body
decay modes of the neutralinos into the sfermion sector. Also
slepton-mediated three-body decays --- which can dominate over the
two-body decays in certain scenarios --- are numerically irrelevant 
for the same reason.

The branching fractions for the relevant decay channels have been
computed at leading order using \FA/\FC\ \cite{Feynarts,FormCalc} and a
purpose-built Fortran code.\footnote{A \FA~model file for the NMSSM has been obtained using {\tt FeynRules} \cite{Christensen:2008py} and {\tt SARAH} \cite{Staub:2009bi}. Details on this implementation will be presented elsewhere.} Results for the neutralino branching ratios in the modified P4 scenario are shown in \reffig{neutdec}.
\begin{figure}
\centering
\includegraphics[width=0.45\columnwidth]{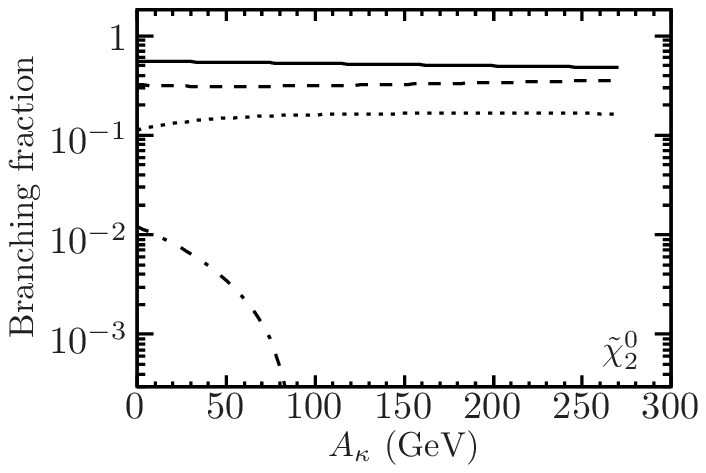}
\includegraphics[width=0.45\columnwidth]{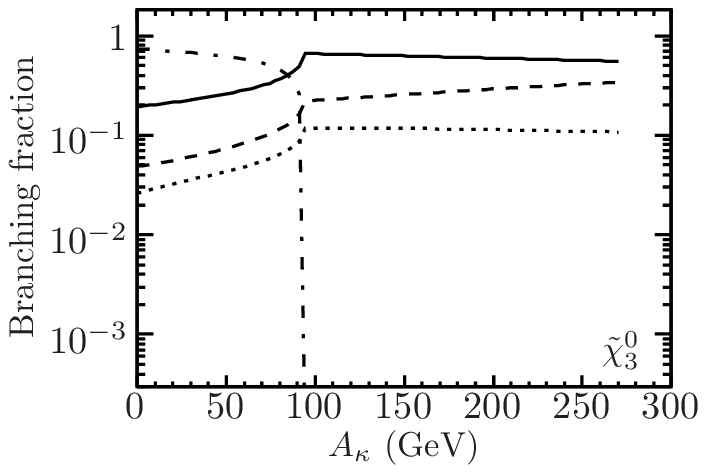}\\
\vspace{3mm}
\includegraphics[width=0.45\columnwidth]{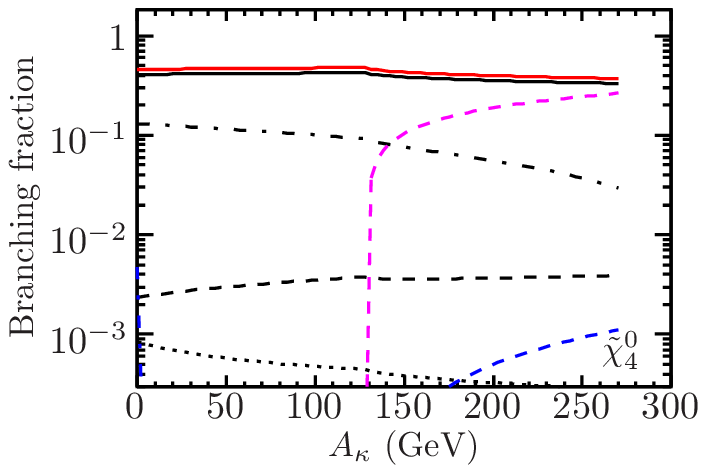}
\includegraphics[width=0.45\columnwidth]{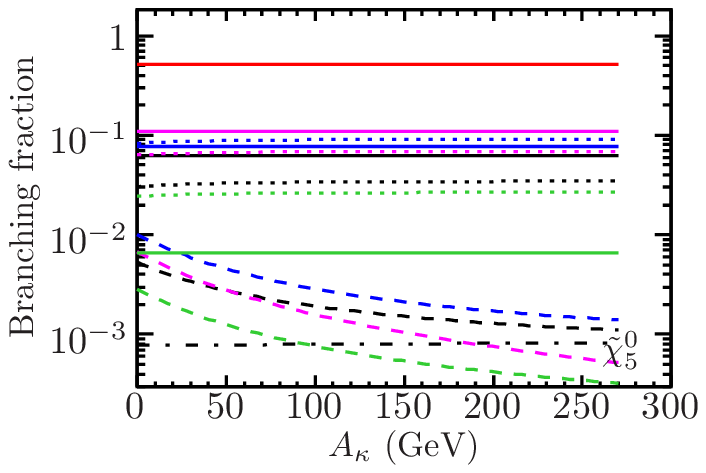}
\caption{Branching ratios in the modified P4 scenario for $\neut{i}\to \neut{j} Z$ (solid), $\neut{i}\to \neut{j} H_1$ (dashed), $\neut{i}\to\neut{j}H_2$ (dotted), and $\neut{i}\to \neut{j}A_1$ (dot-dashed). The color coding indicates the final state neutralino $j=1$ (black), $j=2$ (blue), $j=3$ (magenta), $j=4$ (green), or the chargino mode $\neut{i}\to \cha{1} W^\mp$ (red).}
\label{fig:neutdec}
\end{figure}
The decay modes of $\neut{2}$ and $\neut{3}$ (upper row of
\reffig{neutdec}) --- which are both Higgsino-like --- show similar
patterns for large values of $A_\kappa$. The dominant mode is always
$\neut{i}\to \neut{1}Z$ with a branching ratio of about $50\%$, but the
Higgs channels are also significant with $\mathrm{BR}(\neut{i}\to
\neut{1}H_1)\gtrsim 0.3$ and $\mathrm{BR}(\neut{i}\to \neut{1}H_2)\sim
0.15$. An important  point to note here is that the branching ratios of $\neut{2}$ and
$\neut{3}$ are quite insensitive to changes in $M_{H_1}$ ($A_\kappa$). 
For the heavier neutralinos, $\neut{4}$ and $\neut{5}$ (lower row of
\reffig{neutdec}), which also carry a larger gaugino fraction, the decay
pattern is more complicated. Of largest interest for Higgs production is
the sizable rate for $\neut{4}\to \neut{3}H_1$ (once $A_\kappa$ is
sufficiently large to make this decay mode kinematically possible), and the fact that direct decays of $\neut{5}$ to the LSP are suppressed. This will lead to neutralino decay chains with intermediate (Higgsino) steps. Everything taken together, we can expect a large number of light Higgs bosons to be produced in neutralino cascade decays.

The light chargino $\cha{1}$ decays exclusively into the LSP and a $W$ boson, while the corresponding decay channels for the heavier chargino $\chi_2^\pm$ are shown in \reffig{chardec}.
\begin{figure}
\centering
\includegraphics[width=0.45\columnwidth]{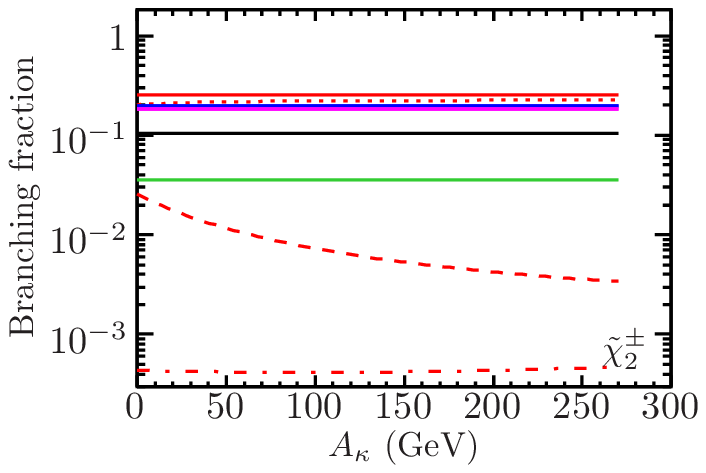}
\caption{Branching ratios of $\cha{2}$ in the modified P4 scenario. Neutralino final states $\cha{2}\to \neut{i} W^\pm$ (solid) are shown with the same color coding as for \reffig{neutdec}. Modes with a final state $\cha{1}$ are shown in red: $\cha{2}\to \cha{1} Z$ (solid), $\cha{2}\to\cha{1}H_1$ (dashed), $\cha{2}\to\cha{1}H_2$ (dotted), $\cha{2}\to \cha{1}A_1$ (dot-dashed).}
\label{fig:chardec}
\end{figure}
Even if the dominant mode is $\cha{2}\to \cha{1}W^\mp$, independently of
$A_\kappa$, there are several channels with a branching fraction of
order $20\%$ of interest for Higgs production. These include the mode
$\cha{2}\to \cha{1}H_2$ and the decays into intermediate-mass Higgsinos,
$\cha{2}\to \neut{2,3}W^\pm$. 

The heavier neutralinos and the heavy chargino, from which a Higgs could
emerge as decay product, can either be produced at the LHC directly or
in the decay of a heavier SUSY particle. The cross section for direct 
production of neutralino pairs is small, only $\mathcal{O}(\mathrm{fb})$ 
at $\sqrt{s}=14\TeV$, and the reach in these channels will be rather
limited even for high luminosity. The large cross sections for production 
of strongly interacting sparticles (squarks and gluinos), on the other
hand, are potentially more promising as a source of the heavier
neutralino states and the heavier chargino. Exploiting cascade decays of
this kind furthermore has the advantage that additional high-$p_T$ jets
are produced, which facilitates triggering and event selection.

We use {\tt Prospino} to calculate the NLO cross sections for production
of $pp\to \tilde{g}\tilde{g}$, $pp\to \tilde{q}\tilde{q}$, $pp\to
\tilde{q}\tilde{\bar{q}}$, and $pp\to \tilde{g}\tilde{q}$ according to
\cite{Beenakker:1996ch}, with CTEQ6\ \cite{Pumplin:2002vw} parton
distributions and a common choice of renormalization and factorization
scales as the average mass of the final state (SUSY) particles.
Numerical results are given in table~\ref{tab:sigma_gg} for the two
centre-of-mass energies $7\TeV$ and $14\TeV$. The cross sections for
$pp\to \tilde{t}\tilde{\bar{t}}$ and $pp\to \tilde{b}\tilde{\bar{b}}$
\cite{Beenakker:1997ut} are also calculated and included in the
analysis, but since they turn out to be significantly smaller than
$\sigma(pp\to \sq\sqb)$ they are not shown in the table. In order to 
give some indication of the expected change in the number of events for
different scenarios, the results are presented for several values of the
squark masses, $M_{\tilde q}$, and the gluino mass, $M_{\tilde g}$. The
mass ranges are selected to respect the published limits from ATLAS
\cite{Aad:2011hh,daCosta:2011qk,Aad:2011ks,Collaboration:2011xk,Aad:2011xm}
and CMS
\cite{Khachatryan:2011tk,Chatrchyan:2011nx,Chatrchyan:2011ah,Chatrchyan:2011bj}
based on the 2010 data. Taking into account also the most recent results 
\cite{ATLAS-CONF-2011-086,CMS-PAS-SUS-11-003},
 the $M_{\tilde{q}}=750\GeV$ case appears to be under some pressure. We present the results of our analysis below for the two 
cases $M_{\mathrm{SUSY}}=750\GeV$ and $M_{\mathrm{SUSY}}=1\TeV$ (the leading order
squark masses are obtained from $M_{\mathrm{SUSY}}$ through \refeq{squarkmassmatrix},
to which higher order corrections are then added).
\begin{table}
\centering
\begin{tabular}{cc|ccccc|ccccc}
\hline
\multicolumn{2}{c}{Masses (GeV)}& \multicolumn{5}{|c|}{$\sigma_{\mathrm{LO}}$ (pb)}  & \multicolumn{5}{|c}{$\sigma_{\mathrm{NLO}}$ (pb)}  \\
$M_{\tilde{g}}$ & $M_{\tilde{q}}$ & $\tilde{g}\tilde{g}$ & $\tilde{q}\tilde{q}$ &$\tilde{g}\tilde{q}$ & $\tilde{q}\tilde{\bar{q}}$ & $\Sigma$ & $\tilde{g}\tilde{g}$ & $\tilde{q}\tilde{q}$ & $\tilde{g}\tilde{q}$& $\tilde{q}\tilde{\bar{q}}$& $\Sigma$\\
\hline
\multicolumn{12}{c}{$\sqrt{s}=7\TeV$} \\
\hline
 $750$ 	& $750$ 	& $0.03$ 	& $0.23$ & $0.25$ & $0.05$ &$0.56$	& $0.07$ & $0.27$ & $0.39$ &$0.08$ &$0.82$\\
 $1000$ 	& $750$ 	& $0.002$ 	& $0.19$ & $0.06$ & $0.05$ &$0.31$	& $0.006$ & $0.21$ & $0.10$ &$0.07$ &$0.39$\\
 $1000$ 	& $1000$ 	& $0.001$ 	& $0.03$& $0.02$ & $0.004$ &$0.06$	& $0.005$ & $0.04$ & $0.04$ &$0.006$ &$0.08$\\
\hline
\multicolumn{12}{c}{$\sqrt{s}=14\TeV$} \\
\hline
$750$ 	& $750$ 	& $1.18$ 	& $1.67$ 	& $5.20$ & $1.06$&$9.11$	& $2.21$ & $2.06$ & $6.78$ &$1.53$ &$12.6$\\
$1000$ 	& $750$ 	& $0.15$ & $1.41$ 	& $1.86$ & $0.96$&$4.38$	& $0.32$ & $1.59$ & $2.44$&$1.34$ &$5.69$\\
$1000$ 	& $1000$ 	& $0.14$ & $0.42$ & $0.87$& $0.18$&$1.61$ & $0.31$ & $0.51$ & $1.19$&$0.26$ &$2.27$ \\
$1500$ 	& $1500$ 	& $0.01$ & $0.04$ & $0.05$& $0.01$&$0.10$ & $0.01$ & $0.05$ & $0.07$&$0.02$ &$0.15$ \\
\hline
\end{tabular}
\caption{Total production cross sections for $pp\to \tilde{g}\tilde{g}$,
$pp\to \tilde{q}\tilde{q}$, $pp\to \tilde{g}\tilde{q}$, and $pp\to
\tilde{q}\tilde{\bar{q}}$ at LO and NLO SUSY-QCD from {\tt Prospino}.
The squark cross sections are summed over the four ``light'' squark
flavours. No kinematic cuts have been applied here. }
\label{tab:sigma_gg}
\end{table}

The nearly mass-degenerate squarks decay preferentially into the SUSY-EW
sector. Direct decays into Higgs bosons (or Higgsinos) are negligible
for squarks of the first two generations due to the small Yukawa
couplings. In contrast to the MSSM, the neutralinos also have a singlino
component to which no squark couples. The left-handed squarks decay
mainly into the wino, $\tilde{q}_L \to \tilde{W}^0 q$, $\tilde{q}_L \to
\tilde{W}^\pm q'$, while the right-handed squarks decay mostly to the
bino, $\tilde{q}_R \to \tilde{B} q$. Numerically, this leads to squark
decay modes listed in table~\ref{tab:squarkdec} for the case with a soft
scalar mass of $M_{\mathrm{SUSY}}=750\GeV$. The squark decay pattern for
$M_{\mathrm{SUSY}}=1\TeV$ is qualitatively similar.\footnote{The main
numerical difference is an increase of $\mathrm{BR}(\tilde{u}_L\to
q'\cha{2})$ to 60\% at the expense of a reduced 
$\mathrm{BR}(\tilde{u}_L\to q'\cha{1})=3.9\times 10^{-2}$.} Since the 
gaugino components are largest in the two heaviest neutralinos, the
neutralinos produced in the squark decays tend to give rise to cascade
decays with several steps.

\begin{table}
\centering
\begin{tabular}{ccccc}
\hline
Final state  & $\tilde{u}_L$ & $\tilde{u}_R$ & $\tilde{d}_L$ &  $\tilde{d}_R$ \\
\hline
$\tilde{q}\to q \neut{1}$ & $2.9\times 10^{-3}$ &$2.6\times 10^{-3}$ &$6.3\times 10^{-3}$ &$2.6\times 10^{-3}$ \\
$\tilde{q}\to q \neut{2}$& $8.1\times 10^{-3}$ &$5.4\times 10^{-3}$ &$1.6\times 10^{-2}$ &$5.4\times 10^{-3}$ \\
$\tilde{q}\to q \neut{3}$& $1.9\times 10^{-3}$ &$4.5\times 10^{-2}$ & $2.0\times 10^{-2}$ & $4.6\times 10^{-2}$\\
$\tilde{q}\to q \neut{4}$& $6.6\times 10^{-2}$ & $0.95$ &$2.9\times 10^{-2}$ & $0.95$\\
$\tilde{q}\to q \neut{5}$& $0.29$ &-- & $0.32$ & --\\
$\tilde{q}\to q' \cha{1}$ & $9.6\times 10^{-2}$ & -- & $3.3\times 10^{-4}$ & --\\
$\tilde{q}\to q' \cha{2}$ &  $0.54$ & -- & 0.61 & --\\
\hline
\end{tabular}

\caption{Branching ratios for the first and second generation squarks into neutralinos and charginos in the modified P4 scenario with $M_{\mathrm{SUSY}}=750\GeV$. Results for channels with a branching ratio below $10^{-4}$ are not shown.}
\label{tab:squarkdec}
\end{table}

Finally, we note that the gluinos decay `democratically' through $\tilde{g}\to \tilde{q}\bar{q}$ into all flavours, with rates governed only by the available phase space.

\section{LHC analysis}
In order to assess whether the process discussed in the previous section can be useful as a
Higgs search channel at the LHC we perform a Monte Carlo simulation. 
Here we use as benchmark the modified P4 scenario with the two different
settings for the soft scalar mass: $M_{\mathrm{SUSY}}=750\GeV$ and
$M_{\mathrm{SUSY}}=1\TeV$. 
The $\overline{\mathrm{DR}}$ value of the gluino mass
parameter is set to $M_{3}=1\TeV$. We select $A_\kappa$ such that $M_{H_1}\simeq 40\GeV$, which also 
affects $M_{H_2}$, $M_{A_1}$ and the branching ratios in the two cases 
as discussed in section \ref{sect:Higgsprod}. We have chosen this
value of $M_{H_1}$ as an illustrative example of our scenario with 
$20\GeV < M_{H_1} < M_Z$ and in order to make
contact with the analyses of the ``CPX hole'' in the MSSM with complex
parameters. Our results however depend only very mildly on the specific 
choice for $M_{H_1}$.
The simulation results are presented below both for LHC running at 
centre-of-mass energies of $7\TeV$ and $14\TeV$.

The squark and gluino-induced cascades in general give rise to a final state with high multiplicities and several hard jets, as well as large missing transverse momentum due to the presence of the LSP at the end of each decay chain. The minimal signal cascades (defined to be those with at least one Higgs boson present) generated by the production of a single squark or gluino correspond to
\begin{align}
\tilde{q}&\to q\neut{i}\to q \neut{1}H_k\to q \neut{1}b\bar{b},\qquad\qquad &n_j\geq 1,\quad n_b\geq 2,\tag{14 a}\label{eq:short}\\
\tilde{g}&\to g\tilde{q}\to gq\neut{i}\to gq \neut{1} H_k\to gq \neut{1}b\bar{b},\qquad\qquad &n_j\geq 2,\quad n_b\geq 2\tag{14 b}
\label{eq:long}.
\end{align}
Equations~\eqref{eq:short} and \eqref{eq:long} show the minimum number
of light and heavy flavour ($b$-) jets expected in the signal. Each
event contains production of a pair of sparticles and their associated
jets, meaning that the full signature for production of at least one
$H_1$ in the hadronic final state will be $n_j \geq 2$, $n_b\geq 2$.
Since direct decays of the heavier (mainly gaugino) neutralinos into the
singlino LSP are practically absent (cf.~\reffig{neutdec}), most signal cascades will contain an intermediate Higgsino step which will add further particles in the final state. The typical jet multiplicity will also be higher due to additional QCD activity, in particular for gluon-initiated processes.

\subsection{Event Generation}
For the event generation, we use {\tt MadGraph/MadEvent~4.4.44} \cite{Alwall:2007st} to calculate the leading order matrix elements for $pp\to \gl\gl,\sq\sq,\gl \sq, \sq\sqb,\st\stb$, and $\tilde{b}\tilde{\bar{b}}$. The different event categories are weighted by the corresponding NLO cross sections to produce an inclusive SUSY sample.
The resonance decay chains are then generated with {\tt PYTHIA 6.4}
\cite{Sjostrand:2006za} using the NMSSM decay rates calculated above as
input through the SUSY Les Houches accord \cite{SLHA2}. The {\tt PYTHIA}
generator is also used to produce additional QCD radiation through 
initial- and final state parton showers, for parton fragmentation, and to generate multiple interactions for the underlying event. This produces fully dressed hadronic events which are passed through the fast simulation of the ATLAS detector performance implemented in the {\tt Delphes} package \cite{Ovyn:2009tx}.\footnote{Running the same {\tt Delphes} analysis with the ``CMS'' detector setup and similar parameters for jets and heavy flavour tagging, no significant differences are observed in the output.} Hadronic jets are clustered using the anti-$k_T$ algorithm \cite{Cacciari:2008gp} with a jet radius measure of $R=0.4$. 

Since for the lightest Higgs boson the decay to $b\bar{b}$ is favored,
the probability $\eta_b$ to correctly identify jets originating from
bottom partons ($b$-tagging efficiency) becomes a crucial quantity for
the analysis. 
Based on \cite{Aad:2009wy} we parametrize this efficiency as a constant $\eta_b=0.6$ with respect to both the detector geometry and the jet energy scale.
Only jets in the central tracking region $|\eta|<2.5$ can be tagged. The
rate for misidentification as a $b$-jet is assumed to be $\eta_c=0.1$
for charm jets, and $\eta_q=0.01$ for jets produced by light quarks and
gluons. The actual tagging algorithm implemented in the {\tt
Delphes}~simulation is not based on a particular experimental method to
identify $b$-jets. The algorithm determines if a jet is close enough in
$\Delta R$ to a ``true'' $b$ parton. When this is the case, the
efficiencies given above are applied to determine if the tagging is
successful or not. 

\subsection{Backgrounds}
Based on the event signature, SM production of $t\bar{t}$ with at least one hadronically decaying $W$ boson (or additional jet activity) constitutes an irreducible background to the Higgs signal. We can \emph{a priori}~expect this to be the most important SM background since the scale for the SUSY-QCD processes is high ($> 1\TeV$). 
In principle there are other sources of background from production of
$W+\mathrm{jets}$ ($b\bar{b}$), $Z+\mathrm{jets}$ ($b\bar{b}$), direct
production of $b\bar{b}+\mathrm{jets}$, or from QCD multijets. The cross
sections for these processes are large compared to the signal cross
section, with QCD multijets the largest and thereby potentially the most
serious. However, for QCD jet production to constitute a background to
the Higgs signal simultaneously a double misidentification of heavy
flavour jets and a large mismeasurement of the missing transverse
energy is required. 
It is furthermore difficult to simulate this background reliably, since
extreme kinematical fluctuations --- or experimental effects --- would
be necessary to produce the signal-like events. A detailed study of the
experimental effects would require a full detector simulation, which is beyond
the scope of the present paper. However, the dominance of the 
$t\bar{t}$ background over other SM processes, such as $W+\mathrm{jets}$
or $Z+\mathrm{jets}$, for our final state has also been demonstrated
experimentally by the results from SUSY searches with $b$-jets and
missing $E_T$ \cite{Aad:2011ks}. We therefore proceed under the
assumption that the cuts devised to suppress the irreducible $t\bar{t}$
background will also be efficient for suppressing the other SM backgrounds 
as well.

For the normalization of the $t\bar{t}$ background we use the NLO cross
section $\sigma(pp\to t\bar{t})=902\,\mathrm{pb}$ ($\sqrt{s}=14\TeV$)
and $\sigma(pp\to t\bar{t})=162\,\mathrm{pb}$ ($\sqrt{s}=7\TeV$), 
computed with the {\tt HATHOR}~package \cite{Aliev:2010zk} for $m_t=173.3\GeV$ and MSTW2008 PDFs \cite{Martin:2009iq}. In this way a consistent NLO normalization is used for both the signal and background events. The $t\bar{t}$ background is generated in the same Monte Carlo framework as already described for the signal.

In addition to the SM backgrounds, the process we are interested in
receives an important background from the SUSY cascade itself. Any final
state containing two $b$-jets which do not result from an intermediate
Higgs boson contributes to this background.  Attempting to suppress the
SUSY background events would require additional cuts that depend on the
kinematics of the decay chains. This is something which may indeed be
possible to devise once information on the supersymmetric spectrum has
become available, but since we do not want to make any particular
assumptions on the pattern of the SUSY spectrum, no selection will be
applied aiming to reduce the SUSY background. Instead we will consider
the inclusive $b\bar{b}$ mass spectrum directly after applying the cuts
designed to reduce the SM background to determine if a Higgs signal can
be extracted. 

\subsection{Event Selection}
As a first step, we perform a preselection of the expected event topology, demanding $n_j \geq 2$, $n_b \geq 2$. All reconstructed jets are required to have a minimum $p_T\geq~25$~GeV. 
\begin{figure}
\centering
\includegraphics[width=0.45\columnwidth]{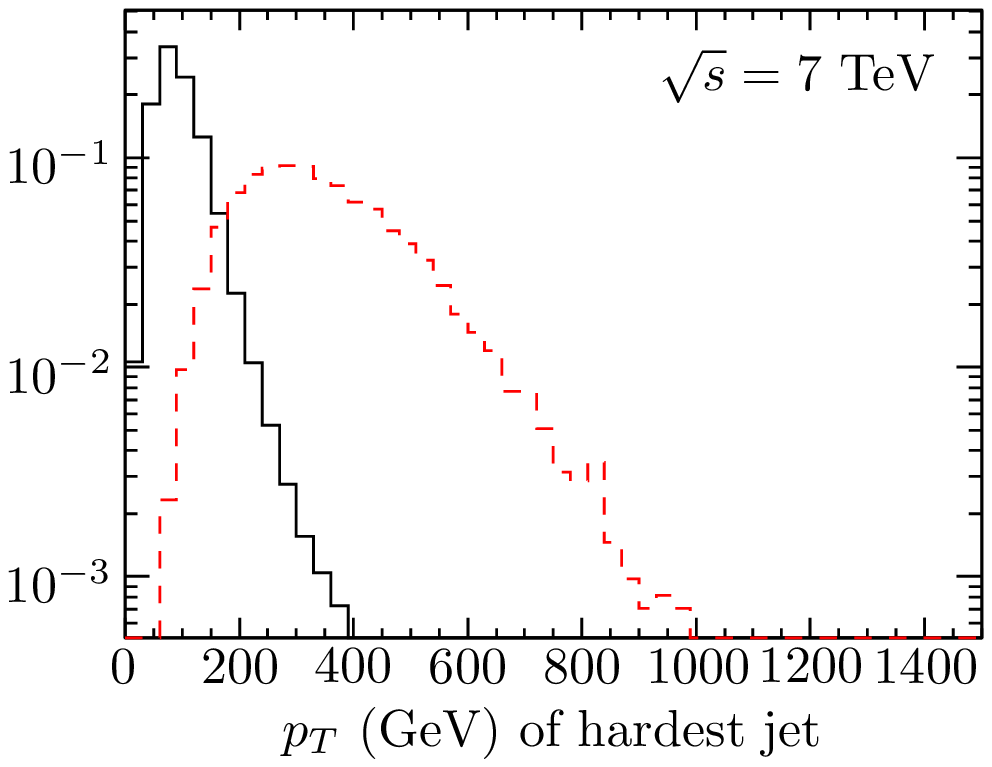}\;
\includegraphics[width=0.45\columnwidth]{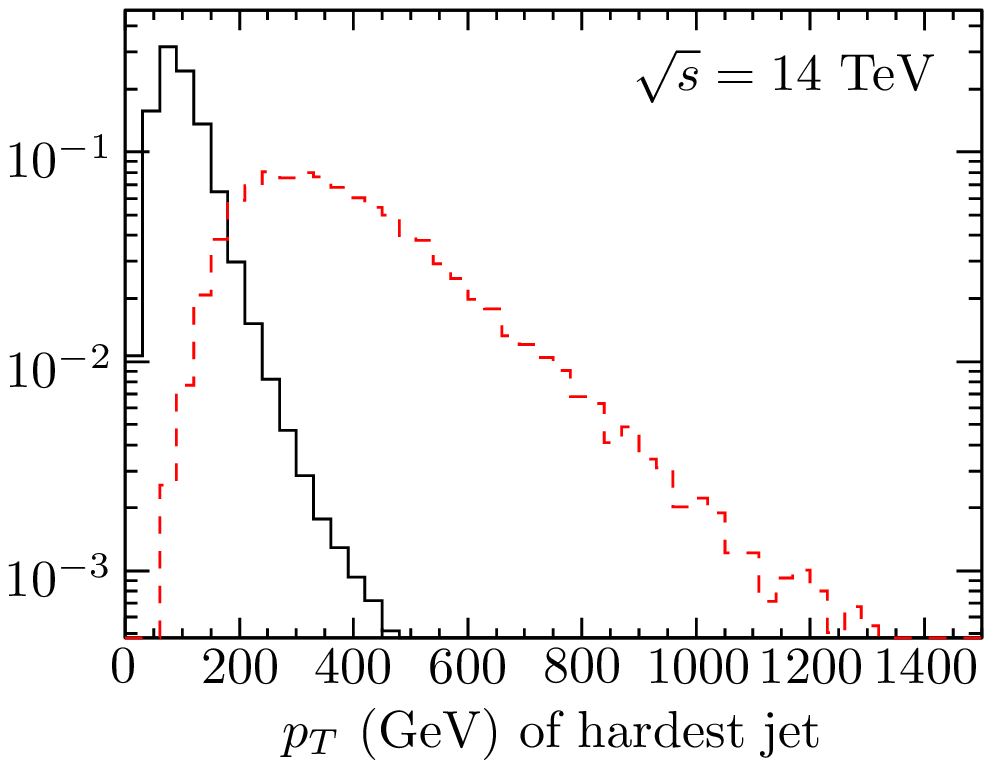}
\caption{Transverse momentum distribution of the hardest (non-$b$) jet at $7$~TeV (left) and $14$~TeV (right) for the inclusive SUSY sample with $M_{\mathrm{SUSY}}=750\GeV$(dashed) and SM $t\bar{t}$ background (solid). The histograms are normalized to unity.}
\label{fig:jet1pt}
\end{figure}
\begin{figure}
\centering
\includegraphics[width=0.45\columnwidth]{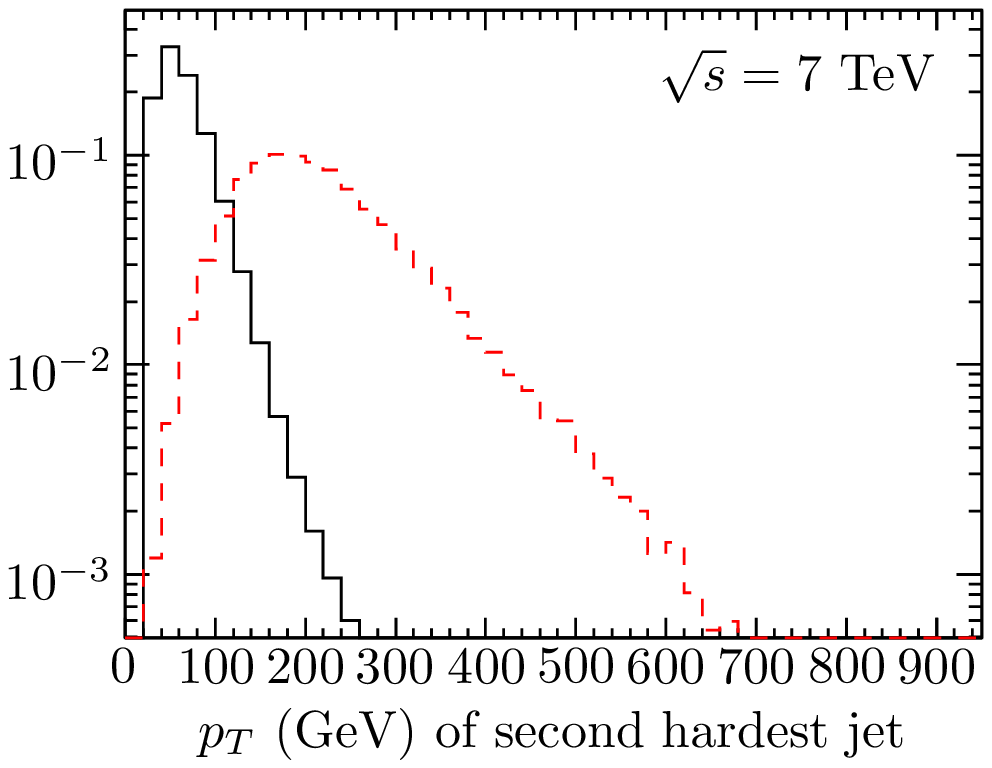}\;
\includegraphics[width=0.45\columnwidth]{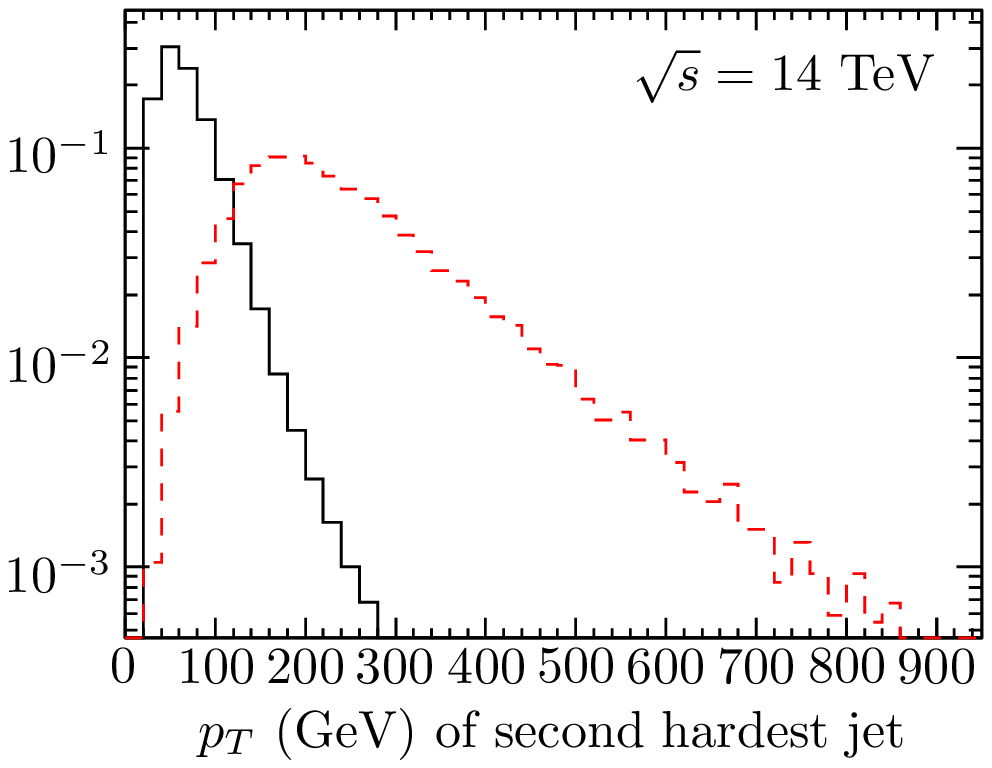}
\caption{Transverse momentum distribution of the second hardest jet at $7$~TeV (left) and $14$~TeV (right) for the inclusive SUSY sample with $M_{\mathrm{SUSY}}=750\GeV$ (dashed) and SM $t\bar{t}$ background (solid). The histograms are normalized to unity.}
\label{fig:jet2pt}
\end{figure}

Figure~\ref{fig:jet1pt} shows the $p_T$ distribution for the hardest jet
in each event, comparing the inclusive SUSY events (with
$M_{\mathrm{SUSY}}=750\GeV$) to the $t\bar{t}$ background. We show the
results for the two cases 
$\sqrt{s}=7\TeV$ (left) and $\sqrt{s}=14\TeV$ (right). In order to
illustrate the effect of applying cuts to this variable, each histogram is normalized to unity.
From figure~\ref{fig:jet1pt} it is clear that the leading jet from the SUSY events has a much harder scale compared to the $t\bar{t}$ events. This can be understood as a result of the large boost obtained by the light quark jets originating from squark decays. It can also be seen that there is only a minor scaling difference in the jet $p_{T}$ distribution between the $7\TeV$ and $14\TeV$ cases. The same is true for the second hardest (light) jet, for which the corresponding $p_{T}$ distribution is shown in \reffig{jet2pt}.
Similar differences between signal and background can be observed also for the third and fourth jet when they are present.

With each cascade ending in the stable LSP, a large missing transverse
energy $\slashed{E}_T$ is expected for the signal events. This
distribution is displayed in \reffig{etmiss}, and shows indeed that the
SUSY distribution peaks at high $\slashed{E}_T$ values ($\gtrsim
200\GeV$). This is therefore an important discriminating variable to
suppress the background from $t\bar{t}$ events, where the missing
transverse energy is due to neutrinos from leptonic $W$ decays. 
As already mentioned, a hard cut on $\slashed{E}_T$ is also necessary to
suppress the background from ordinary QCD multijet events and direct production of $b\bar{b}$. A further advantage of the large $\slashed{E}_T$ is that it can be used for triggering.
\begin{figure}
\centering
\includegraphics[width=0.45\columnwidth]{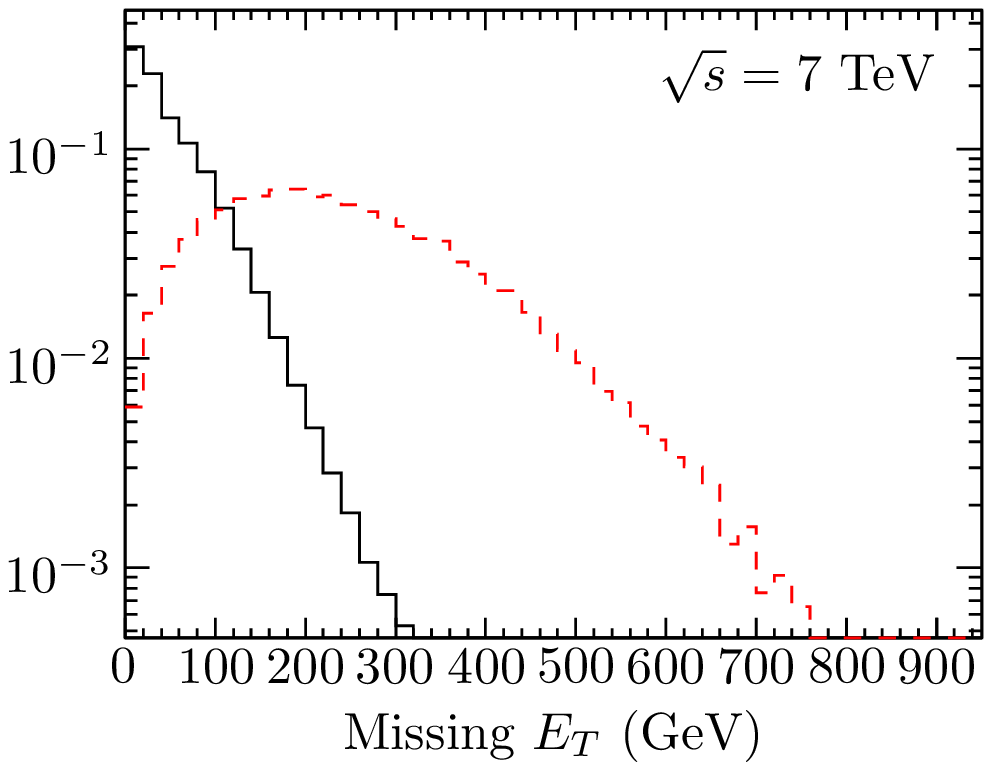}\;
\includegraphics[width=0.45\columnwidth]{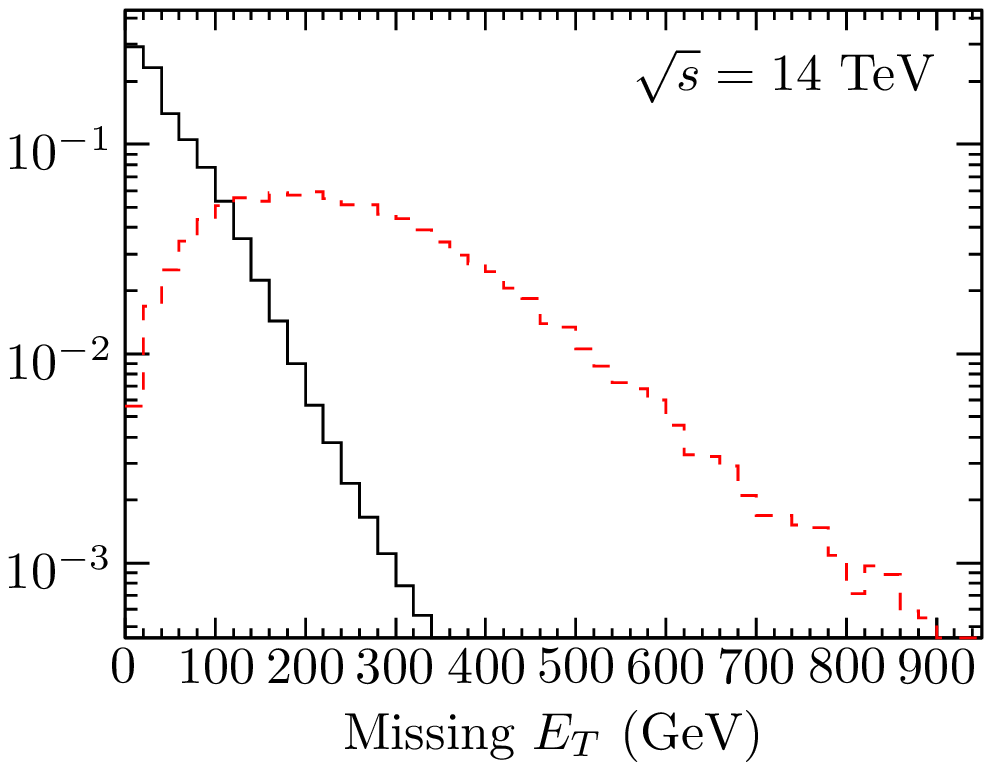}
\caption{Missing transverse energy at $7$~TeV (left) and $14$~TeV (right) for the inclusive SUSY sample (dashed) and SM $t\bar{t}$ background (solid). The histograms are normalized to unity.}
\label{fig:etmiss}
\end{figure}

The final kinematical distribution we are going to consider is displayed in \reffig{bbdr}. It shows the separation in $\Delta R=\sqrt{(\Delta \eta)^2+(\Delta \phi)^2}$ between pairs of $b$-jets. For events with $n_b> 2$ all possible combinations have been included. The signal distribution is seen to peak near the minimum separation of $\Delta R=0.4$ set by the jet measure, while the $t\bar{t}$ background prefers the $b$-jets to be more back-to-back and peaks at $\Delta R\sim \pi$.
\begin{figure}
\centering
\includegraphics[width=0.45\columnwidth]{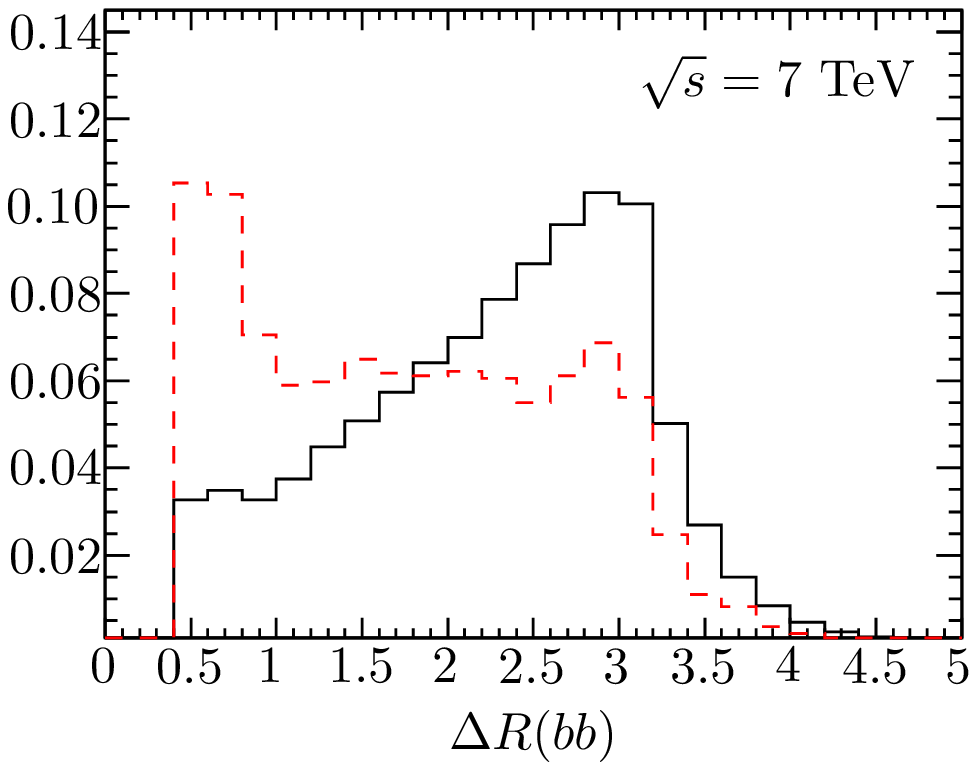}\;
\includegraphics[width=0.45\columnwidth]{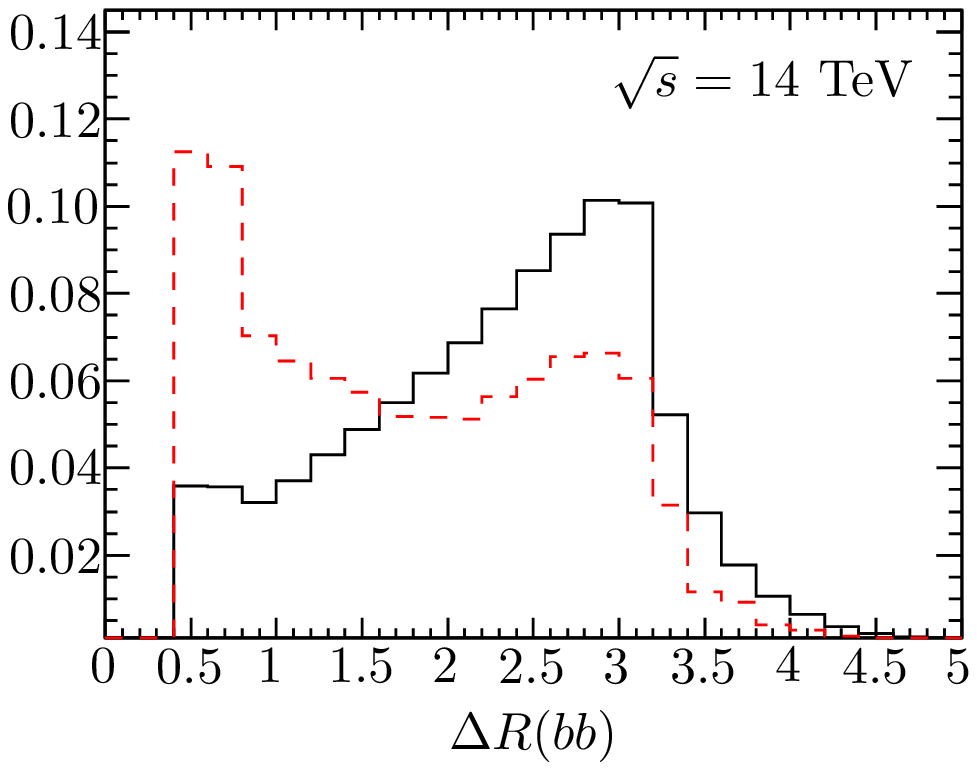}
\caption{Distribution in $\Delta R(b\bar{b})$ at $7$~TeV (left) and $14$~TeV (right) for the inclusive SUSY sample (dashed) and SM $t\bar{t}$ background (solid). The histograms are normalized to unity. In events containing more than two $b$-jets, all possible combinations have been included.}
\label{fig:bbdr}
\end{figure}

\begin{table}
\centering
\begin{tabular*}{0.9\textwidth}{@{\extracolsep{\fill}} lrrr}
$M_{\mathrm{SUSY}}=750$~GeV  & Incl.~SUSY & Signal & Background \\
\hline
Generated events & 18 465 & 8 261 & 10 195 \\
$n_j\geq 2$, $n_b\geq 2$& 4 048 & 2 786 & 1 262  \\
$p_T^{\mathrm{jet 1}} > 250\GeV$, $p_T^{\mathrm{jet 2}} > 100\GeV$ & 2 436 & 1 738 & 698 \\
$\slashed{E}_T > 150\GeV$ & 1 735 & 1 211 & 524 \\
$\mathrm{min}\left[\Delta R(bb)\right]<1.5$ & 1 014 & 774 & 240 \\
\hline
Total efficiency & $5.5\times 10^{-2}$ & $9.4\times 10^{-2}$ & $2.4\times 10^{-2}$ \\
\hline
\\
$M_{\mathrm{SUSY}}=1$~TeV & Incl.~SUSY & Signal & Background \\
\hline
Generated events & 20 671 & 10 923 & 9 748 \\
$n_j\geq 2$, $n_b\geq 2$ & 5 313 & 4 344 & 969   \\
$p_T^{\mathrm{jet 1}} > 250\GeV$, $p_T^{\mathrm{jet 2}} > 100\GeV$ & 4 642 & 3 828 & 814 \\
$\slashed{E}_T > 150\GeV$ & 3 705 & 3 036 & 669  \\
$\mathrm{min}\left[\Delta R(bb)\right]<1.5$ & 2 544 & 2 170 & 374 \\
\hline
Total efficiency & $0.12$ & $0.20$ & $3.8\times 10^{-2}$ \\
\hline
\end{tabular*}
\caption{Number of events remaining after each step of the event selection at $\sqrt{s}=7 \TeV$. The SUSY events are classified as signal or background based on the presence of (at least one) Higgs boson in the decay chain. The total number of generated events in the inclusive sample is arbitrary.}
\label{tab:cuts7}
\end{table}
The precise cuts applied --- and their effect on the event selection ---
are shown for the SUSY events in table~\ref{tab:cuts7} (for the 7 TeV
case) and table~\ref{tab:cuts14} (14 TeV). Table~\ref{tab:cutsttbar}
gives the corresponding information for the SM $t\bar{t}$ background. Note that the number of generated events in these tables does not correspond to any particular luminosity, but is rather selected to give adequate statistics for the event selection. The inclusive SUSY sample is split into signal and background, where the signal consists of the events containing at least one Higgs boson (as determined from Monte Carlo truth information). In the last row we give the accumulated total efficiencies of all the cuts.
Looking first at table~\ref{tab:cuts7}, we see that an efficiency of
$5.5\times 10^{-2}$ is obtained for the case with
$M_{\mathrm{SUSY}}=750\GeV$. This efficiency is more than doubled (0.12)
for the case with $M_{\mathrm{SUSY}}=1\TeV$, since the heavier squarks
give harder jets as decay products which leads to more events passing
the jet $p_T$ cuts. The larger boost given to the LSP at the end of the
decay chain also leads to an increased $\slashed{E}_T$. The same
qualitative features are visible at $14\TeV$, as can be read off
table~\ref{tab:cuts14}. Due to the favorable signal statistics at
$14\TeV$,\footnote{Going from $7\TeV$ to $14\TeV$, the signal cross
section for $M_{\mathrm{SUSY}}=750\GeV$ increases by a factor $14.5$,
while the $t\bar{t}$ cross section is only increased by a factor $5$.}
we can afford slightly harder cuts on $\slashed{E}_T$ and $\Delta R(bb)$
in this case, something which is also needed to maintain a good
background suppression. One should therefore not be discouraged by the
somewhat lower efficiencies recorded in this case ($4.2\times 10^{-2}$
for $M_{\mathrm{SUSY}}=750\GeV$ vs.\ $9.8\times 10^{-2}$ for $M_{\mathrm{SUSY}}=1\TeV$). The signal efficiencies can be compared to those for the $t\bar{t}$ background, given in table~\ref{tab:cutsttbar}, which are at the $10^{-5}$ level for both energies. It is clear from this table that the hard cuts on the jet $p_T$ and the $\slashed{E}_T$ distribution are the most important handles available to suppress the background.
\begin{table}
\centering
\begin{tabular*}{0.9\textwidth}{@{\extracolsep{\fill}} lrrr}
$M_{\mathrm{SUSY}}=750$~GeV  & Incl.~SUSY & Signal & Background \\
\hline
Generated events & 23 771 & 10 874 & 12 897\\ 
$n_j\geq 2$, $n_b\geq 2$ 					 & 5 009 & 3 610 & 1 399  \\
$p_T^{\mathrm{jet 1}} > 250\GeV$, $p_T^{\mathrm{jet 2}} > 100\GeV$ & 3 287 & 2 422 & 865 	 \\
$\slashed{E}_T > 200\GeV$ 					   & 1 935 & 1 400 & 535   \\
$\mathrm{min}\left[\Delta R(bb)\right]<1.2$  & 991 	& 775   & 216   \\
\hline
Total efficiency & $4.2\times 10^{-2}$ & $7.1\times 10^{-2}$ & $1.7\times 10^{-2}$ \\
\hline
\\
$M_{\mathrm{SUSY}}=1$~TeV  & Incl.~SUSY & Signal & Background    \\
\hline
Generated events & 20 232 & 10 557 & 9 675 \\
$n_j\geq 2$, $n_b\geq 2$ & 5 428 & 4 338 & 1 090   \\
$p_T^{\mathrm{jet 1}} > 250\GeV$, $p_T^{\mathrm{jet 2}} > 100\GeV$ & 4 852 & 3 924 & 928   \\
$\slashed{E}_T > 200\GeV$ & 3 392 & 2 719 & 673  \\
$\mathrm{min}\left[\Delta R(bb)\right]<1.2$ & 1 983 & 1 673 & 310  \\
\hline
Total efficiency & $9.8\times 10^{-2}$ & $0.16$ & $3.2\times 10^{-2}$ \\
\hline
\end{tabular*}
\caption{Events remaining after each step of the event selection at $\sqrt{s}=14 \TeV$. The event categories are similar to those in table~\ref{tab:cuts7}. The total number of generated events in the inclusive sample is arbitrary.}
\label{tab:cuts14}
\end{table}
\begin{table}
\centering
\begin{tabular}{lrr}
 SM $t\bar{t}$ background & 7 TeV & 14 TeV\\
\hline
Generated events & 900 000 & 2 000 000 \\
$n_j\geq 2$, $n_b\geq 2$ & 259 110& 576 232\\
$p_T^{\mathrm{jet 1}} > 250\GeV$, $p_T^{\mathrm{jet 2}} > 100\GeV$ & 1 120 & 5 189\\
$\slashed{E}_T > 150 \GeV$ (7 TeV), $> 200 \GeV$ (14 TeV) & 102 & 405\\
$\mathrm{min}\left[\Delta R(bb)\right]<1.5$ (7 TeV), $<1.2$ (14 TeV) & 12 & 61\\
\hline
Total efficiency & $1.3\times 10^{-5}$ & $3.0\times 10^{-5}$ \\
\hline
\end{tabular}
\caption{SM $t\bar{t}$ background events remaining after each step of the event selection at $\sqrt{s}=7 \TeV$ and $\sqrt{s}=14 \TeV$. The total number of generated events is arbitrary.}
\label{tab:cutsttbar}
\end{table}

As discussed in the previous section, we do not apply any specific cuts
to suppress the background from SUSY events that do not involve a Higgs
boson. The numbers given in tables \ref{tab:cuts7} and 
\ref{tab:cuts14} show that nevertheless our event selection gives rise
to an improvement also in the ratio of signal events over
SUSY-background events. 
The largest difference in selection efficiency between the SUSY signal
and background arises from the typical number of $b$ quarks produced in the two cases, which is larger for the events where Higgs bosons are produced, leading to  a stronger reduction of the SUSY background by the jet multiplicity cut. The cut on $\Delta R$ also contributes to the difference. This cut has the pleasant ``side effect'' to enrich the SUSY sample in Higgs events since the jets resulting from $H_1\to b\bar{b}$ decays are more likely to show up for small $\Delta R$ than those from two unpaired $b$-jets.

\subsection{Results}
Figure~\ref{fig:bbmass7} shows the resulting $b\bar{b}$ mass spectra after final event selection for an integrated luminosity of $5$ fb$^{-1}$ at $7\TeV$. For events with $n_b> 2$ only the $b$-jet combination minimizing $\Delta R(bb)$ has been included. This reduces effects of combinatorics and increases the 
sensitivity for discovering resonances in the low mass region. For the
scenario with relatively
light squarks ($M_{\mathrm{SUSY}}=750\GeV$) shown in the left plot, we observe two peaks close to the masses of the $Z$ boson and $H_1$, respectively. There is also a continuous distribution with a tail towards much higher values for 
$M_{b \bar b}$. This results from false pairings, fake $b$-jets, or from $b$-jet pairs of non-resonant origin such as $t$ or $\tilde{b}$ decays.  
The same qualitative features are visible in the signal for $M_{\mathrm{SUSY}}=1\TeV$ (right plot), but the statistics is 
rather poor due to the low signal cross section.
In \reffig{bbmass14} we show the
$M_{b\bar{b}}$ distribution at $14\TeV$, again for an integrated
luminosity of $5$ fb$^{-1}$. Here the signal statistics is much higher,
so that a clear distinction of the $H_1$ resonance from the background
should be possible 
both for $M_{\mathrm{SUSY}}=750\GeV$ (left) and 
$M_{\mathrm{SUSY}}=1\TeV$ (right).

The same distributions are shown in figures \ref{fig:bbmass_stack7} (for
the LHC at 7 TeV) and \ref{fig:bbmass_stack14} (for 14 TeV), but here with 
stacked histograms to more closely resemble ``real'' data. 
Here we have furthermore split up the inclusive SUSY sample into signal
events (displayed in red), characterised by the presence of (at least) one Higgs boson in
the decay chain, and the remaining SUSY background
events (black). The latter constitutes an additional source of
background besides the SM $t\bar{t}$ background (light gray).
In \reffig{bbmass_stack7} we see that the most striking feature is the 
$H_1$ peak. Although the $t\bar{t}$ background peaks at roughly the same 
position as the signal, the statistics of signal events should be
sufficient for establishing a signal over the background.
In the 14 TeV case, figure \ref{fig:bbmass_stack14} illustrates the
features observed already in \reffig{bbmass14}. The $H_1$ peak stands out
clearly above the background distribution, both for 
$M_{\mathrm{SUSY}}=750\GeV$ and $M_{\mathrm{SUSY}}=1\TeV$.

\begin{figure}
\centering
\includegraphics[width=0.48\columnwidth]{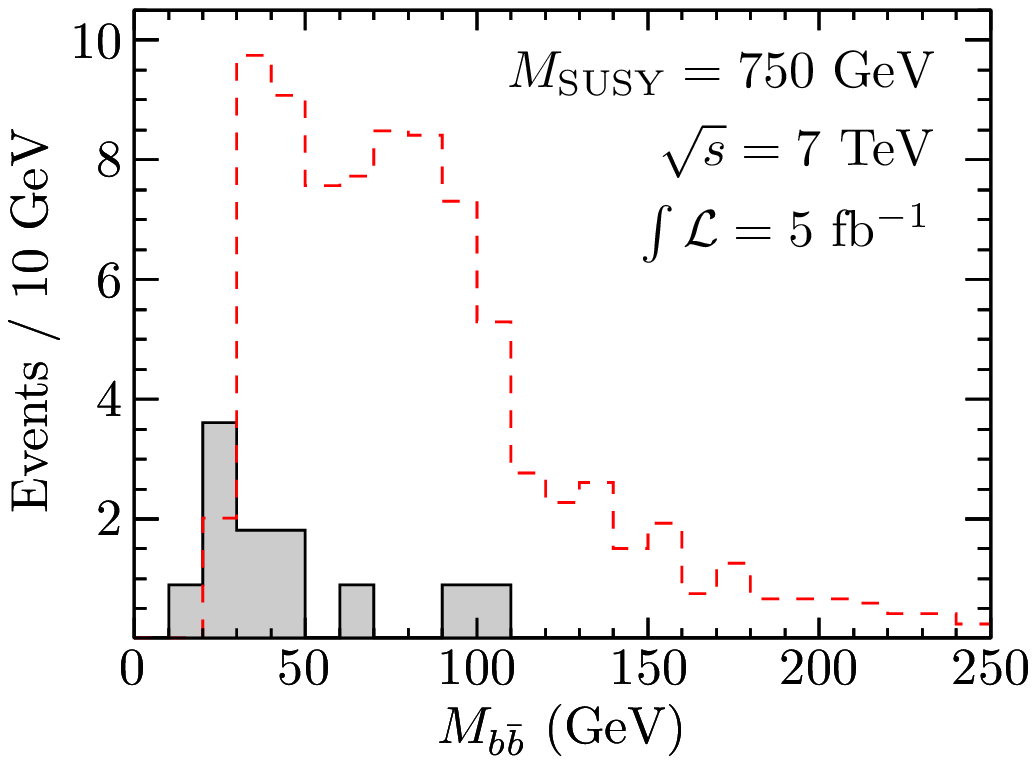}\; \;
\includegraphics[width=0.48\columnwidth]{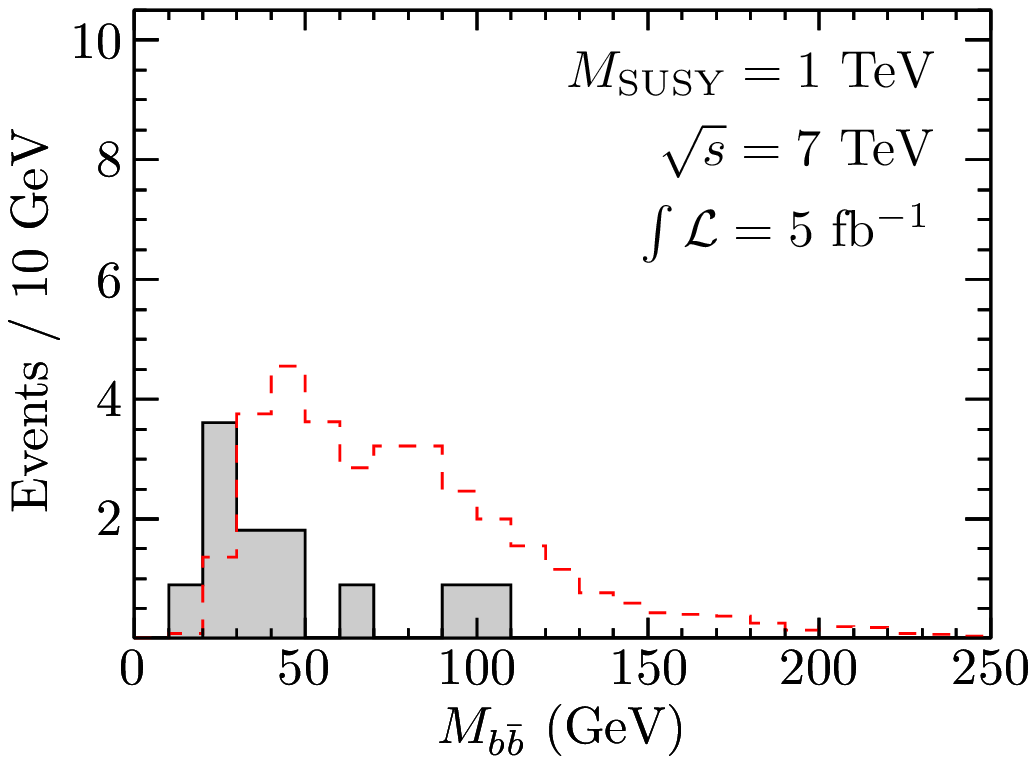}
\caption{Invariant mass of $b$-jet pairs after final event selection for the inclusive SUSY sample (red, dashed) and SM $t\bar{t}$ background (solid) at $7$~TeV in the modified P4 scenario with $M_{\mathrm{SUSY}}=750\GeV$ (left) and $M_{\mathrm{SUSY}}=1\TeV$ (right). The histograms have been normalized to an integrated luminosity of $5$ fb$^{-1}$.}
\label{fig:bbmass7}
\end{figure}
\begin{figure}
\centering
\includegraphics[width=0.48\columnwidth]{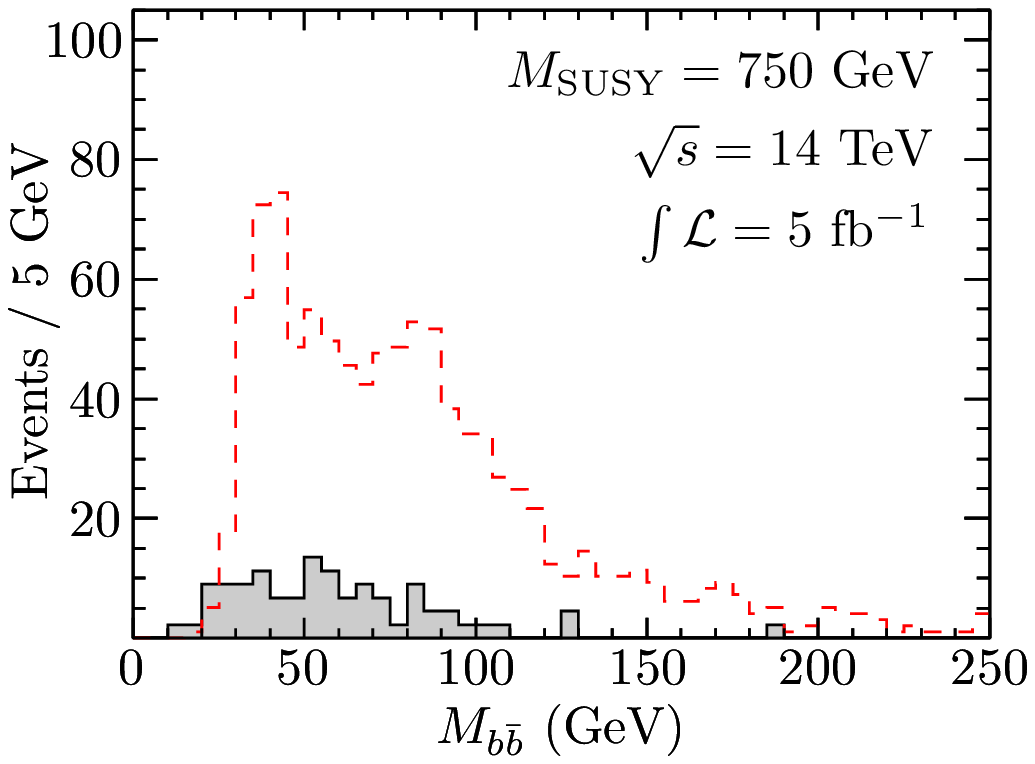}\; \;
\includegraphics[width=0.48\columnwidth]{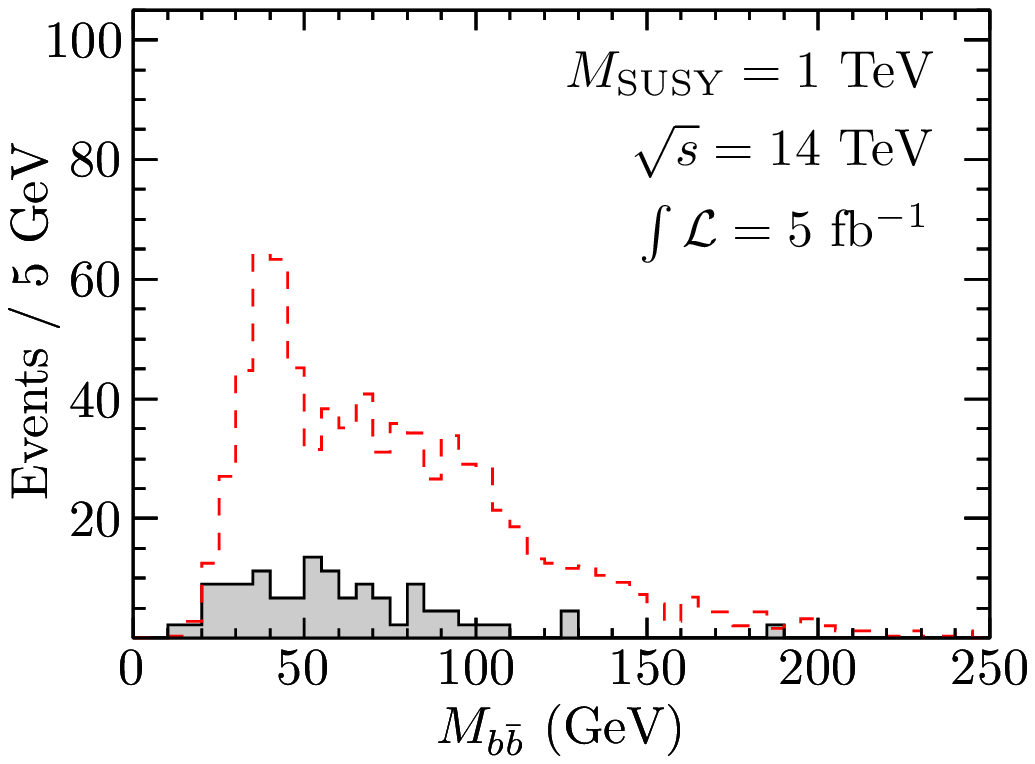}
\caption{Invariant mass of $b$-jet pairs after final event selection for the inclusive SUSY sample (red, dashed) and SM $t\bar{t}$ background (solid) at $14$~TeV in the modified P4 scenario with $M_{\mathrm{SUSY}}=750\GeV$ (left) and $M_{\mathrm{SUSY}}=1\TeV$ (right). The histograms have been normalized to an integrated luminosity of $5$ fb$^{-1}$.}
\label{fig:bbmass14}
\end{figure}

With the assumed statistics, no peaks are observed in the $b\bar{b}$ mass spectrum for the heavier Higgs bosons $H_2$ and $A_1$. This is mainly due to the smallness of the branching ratios into the $b\bar{b}$ mode because of the open Higgs decay channels. Part of the difficulty in observing the heavier resonances is also a result of selecting the combination minimizing $\Delta R(bb)$ in configurations with multiple $b$-jets, which favors selection of the light $H_1$.

\begin{figure}
\centering
\includegraphics[width=0.48\columnwidth]{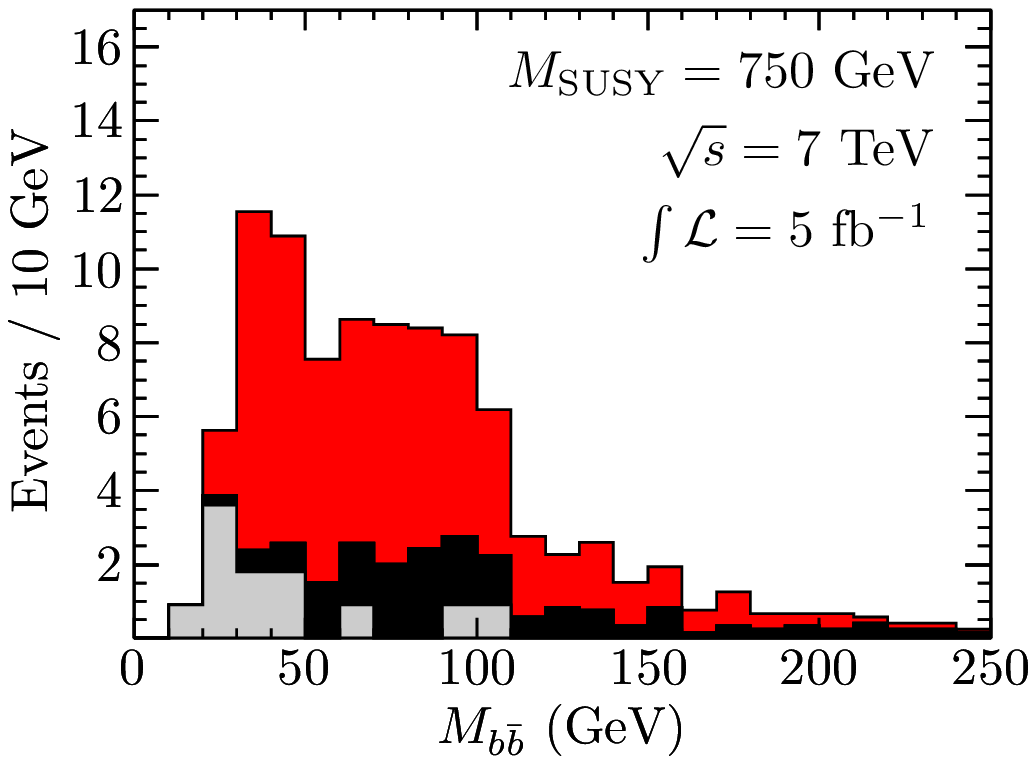}\; \;
\includegraphics[width=0.48\columnwidth]{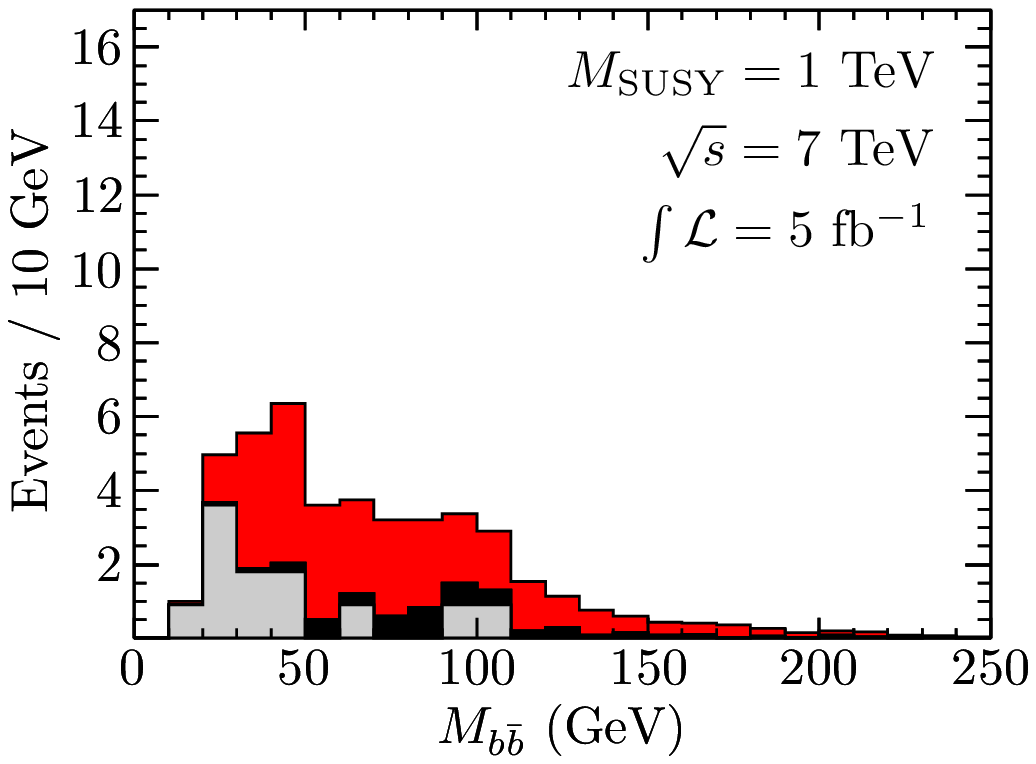}
\caption{Invariant mass of $b$-jet pairs for SUSY signal (red), SUSY background (black) and SM $t\bar{t}$ background (light gray) in the modified P4 scenario with $M_{\mathrm{SUSY}}=750\GeV$ (left) and $M_{\mathrm{SUSY}}=1\TeV$ (right) at $7$~TeV for an integrated luminosity of $5$ fb$^{-1}$.}
\label{fig:bbmass_stack7}
\end{figure}
\begin{figure}
\centering
\includegraphics[width=0.48\columnwidth]{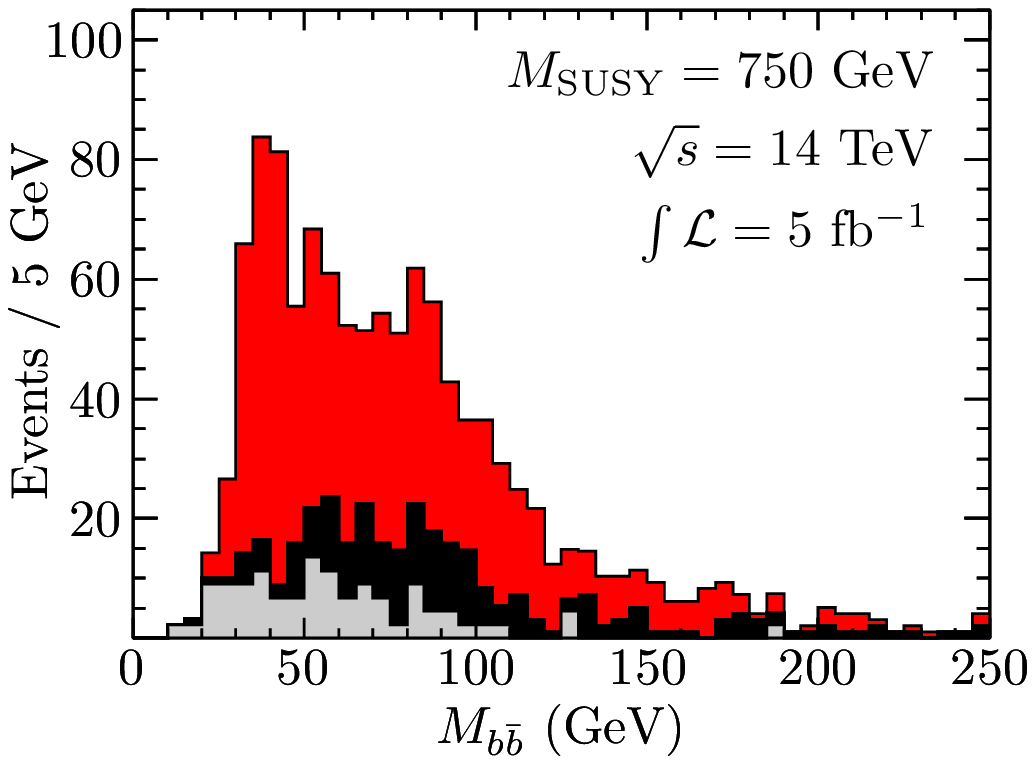}\; \;
\includegraphics[width=0.48\columnwidth]{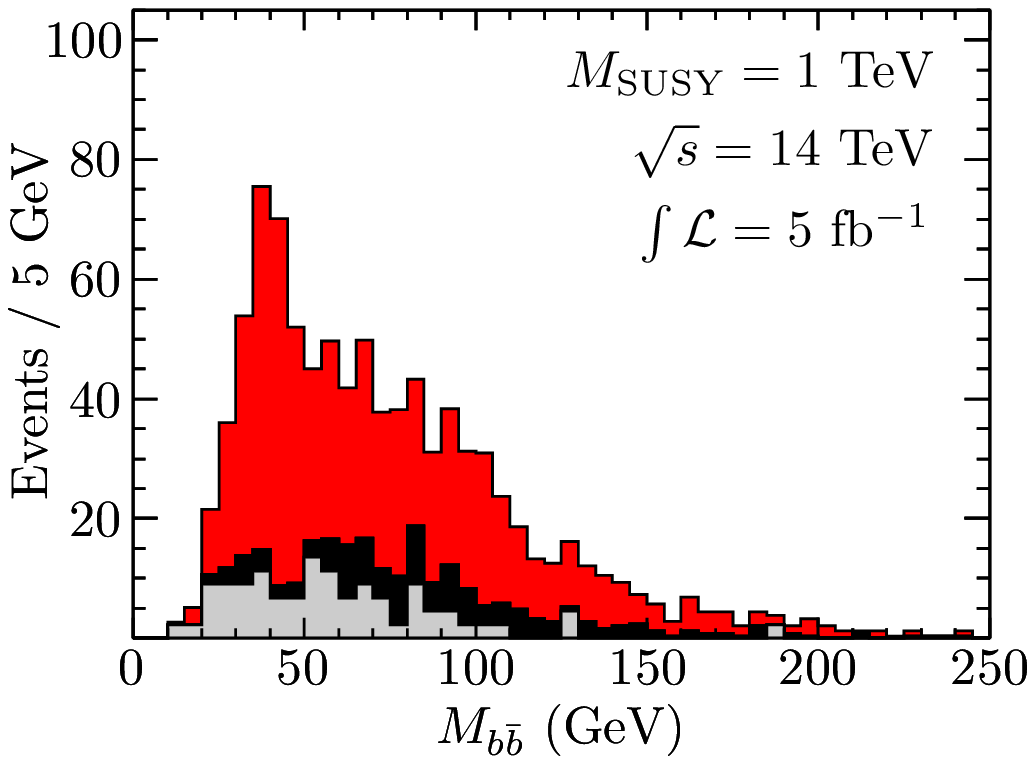}
\caption{Invariant mass of $b$-jet pairs for SUSY signal (red), SUSY background (black) and SM $t\bar{t}$ background (light gray) in the modified P4 scenario with $M_{\mathrm{SUSY}}=750\GeV$ (left) and $M_{\mathrm{SUSY}}=1\TeV$ (right) at $14$~TeV for an integrated luminosity of $5$ fb$^{-1}$.}
\label{fig:bbmass_stack14}
\end{figure}
In order to obtain an estimate of the significance of the $H_1$ mass peak 
we have performed a Gaussian fit to the maximum of the distributions in 
figures \ref{fig:bbmass_stack7} and \ref{fig:bbmass_stack14}.
Table~\ref{tab:sign} lists the results extracted from the
fit for the mean value $M_H$ and the $1\,\sigma$ width $\Delta M_H$ of the Gaussian peak.
 We find that the fitted central values reproduce 
well the correct $H_1$ mass for all cases (recall that the input mass
used in our numerical simulation is $M_{H_1}=40\GeV$). The statistical 
uncertainty on the mean value $M_H$ from the fit is about $\pm 3\GeV$ 
for the LHC energy of $7\TeV$ and about $\pm 1\GeV$ for $14\TeV$. 
This reflects the lower signal statistics available in the low energy 
running, and the more coarse binning in $M_{b \bar b}$ required to observe 
the peak.

The number of signal and background events
in the peak region is obtained by integrating $M_{bb}$ over the interval $[M_H-\Delta M_H, M_H+\Delta M_H]$, corresponding to $\pm 1\,\sigma$ of the Gaussian distribution.
As explained above, the combined background includes both the events from SM $t\bar{t}$ and the part of the inclusive SUSY sample containing no
Higgs bosons in the cascades. The event numbers are combined into the
ratios of signal/background ($S/B$) and $S/\sqrt{B}$ given in
table~\ref{tab:sign}.
We use $S/\sqrt{B}$ as a simple illustration for the expected
significance and
in particular for comparing between the four example cases we consider
here and with other theoretical studies using the same criterion.
Clearly, claiming an actual discovery would require a more sophisticated
statistical treatment. We regard it nevertheless as encouraging that
a significance of $S/\sqrt{B} > 5$ is achieved for 
three of the four cases considered in table \ref{tab:sign}.
The only exception is the case $M_{\mathrm{SUSY}}=1\TeV$ for the LHC at
$7\TeV$, where $S/\sqrt{B} = 3.9$ with $5$ fb$^{-1}$. A somewhat higher
luminosity (or a combination of ATLAS and CMS data) would be needed in
this case in order to reach a significance $S/\sqrt{B} \geq 5$.
The very high significances of about 30 obtained for the $14\TeV$ case
illustrate the qualitative features already observed in the discussion
of figures \ref{fig:bbmass14} and \ref{fig:bbmass_stack14}: there should
be no problems in establishing a signal in this mass region.
\begin{table}
\centering
\begin{tabular*}{0.8\textwidth}{@{\extracolsep{\fill}} cccccc}
\hline
$\sqrt{s}$ & $M_{\mathrm{SUSY}}$~(GeV) & $M_{H}$ (GeV) & $\Delta M_H$ (GeV) & $S/B$ & $S/\sqrt{B}$  \\
\hline
7 TeV & $750$ & $41.6$ & $12.2$ & $3.4$ & $8.2$\\
7 TeV & $1000$ & $37.7$ &$17.6$ &$1.4$ & $3.9$ \\
14 TeV & $750$ & $39.5$ & $8.0$ &$4.5$ & $29.7$\\ 
14 TeV & $1000$ & $39.4$ & $9.7$ & $4.3$ & $29.3$\\ 
\hline
\end{tabular*}
\caption{Mean value $M_H$ and width $\Delta M_H$ extracted from a Gaussian fit to the 
$b \bar b$ mass peak. The corresponding number of signal ($S$) and background ($B$) events is recorded in the $\pm \Delta M_{H}$ interval around the fitted resonance mass. All results are presented for an integrated luminosity of $5$ fb$^{-1}$.}
\label{tab:sign}
\end{table}

\section{Summary and Conclusions}
The NMSSM is both theoretically appealing as an extension of the SM and 
interesting phenomenologically, as its spectrum may contain Higgs bosons 
with mass much below the limits in the SM or the MSSM. We have investigated an
NMSSM scenario with a light CP-even Higgs in the mass range 
$20\GeV < M_{H_1} < M_Z$. Scenarios like this may be missed with the 
standard Higgs search channels at the LHC, in particular due to a
potentially large branching ratio of the heavier $H_2$ state, that has
SM-like couplings to gauge bosons, into a pair of light Higgses. We have
pointed out that there are good prospects for discovering such a light
Higgs boson in SUSY cascade decays at the LHC.

We have performed a Monte Carlo simulation of the signal and the
dominant background to the level of fast detector simulation, taking 
into account also background from other SUSY
events that do not involve cascade decays containing a Higgs boson. 
For our numerical analysis we adapted the ``P4'' benchmark point
proposed for the NMSSM, choosing $M_{H_1} = 40\GeV$ as example value 
for the mass of the light Higgs. 
Production of squarks and gluinos via the strong interaction at the
LHC may give rise to cascade decays involving heavy neutralinos 
and charginos decaying into lighter ones and a light Higgs.
We have investigated the impact of various kinematical variables on
discriminating between the inclusive SUSY signal (including events both with and without a
Higgs boson in the cascade) and the SM background
from $t \bar t$ production.  A set of simple cuts 
has been devised that turned out to be efficient for establishing the
inclusive SUSY signal. 
We did not assume any specific knowledge about the background from SUSY
events without a Higgs in the cascades. Accordingly,
besides favoring events containing the light $H_1$ by selecting the combination minimizing $\Delta R(bb)$ in configurations with multiple $b$-jets, we have not applied any particular cuts for suppressing the SUSY background. 

Our results show that reconstruction of the decay of the light Higgs
into $b \bar b$  may be feasible.
 Such an observation would be a direct experimental sign of the
bottom Yukawa coupling, which is difficult to access in standard search
channels. We have investigated two values of the soft SUSY-breaking
parameter in the squark sector, $M_{\mathrm{SUSY}}=750\GeV$ and
$M_{\mathrm{SUSY}}=1\TeV$, while we set the gluino mass parameter to $1\TeV$.
A modest integrated luminosity of $5$ fb$^{-1}$ has been considered for LHC
running both at $7\TeV$ and $14\TeV$. We find a statistical significance for 
the $H_1$ mass peak of $S/\sqrt{B} \approx 4$ for $M_{\mathrm{SUSY}}=1\TeV$ 
at $\sqrt{s}=7\TeV$. 
This significance increases to $S/\sqrt{B} \approx 8$ for 
$M_{\mathrm{SUSY}}=750\GeV$ at $7\TeV$ and reaches a level of almost 30 
 for both values of $M_{\mathrm{SUSY}}$ at $14\TeV$. While the example
values that we have chosen for $M_{\mathrm{SUSY}}$ and the gluino mass
are close to the current search limits from the LHC, the large
statistical significance that we have found for the $14\TeV$ case
indicates that there is certainly scope to extend our analysis 
to scenarios with heavier squarks and gluinos or to scenarios with reduced 
branching ratios of the neutralinos into Higgs bosons. Since the high-energy 
run of the LHC is not imminent, we leave a more detailed analysis of this reach for future work.

The results presented here have been obtained in a specific
benchmark scenario, but it is easy to see that they are more generally
applicable. First of all, the value $M_{H_1}=40\GeV$ used in our numerical 
analysis was chosen just for illustration.
Our results are rather 
insensitive to the precise value of $M_{H_1}$. Since the production relies 
on the decay of heavier SUSY states, with branching ratios largely independent 
of $M_{H_1}$, the Higgs production rates remain similar for the whole mass 
range $M_{H_1}<M_Z$. The event selection and signal identification through 
$H_1\to b\bar{b}$ proceeds along similar lines as we have discussed.

Concerning the settings of the other SUSY parameters, our results will
be similar for other scenarios fulfilling a few simple criteria:
Obviously, the neutralinos and charginos have to be sufficiently lighter
than the squarks and gluinos in order to be produced at all in the
cascade decays of the latter. The squark decays also provide the hard
jets utilized in the event selection. With the present limits from the
LHC searches on the masses of the gluino and the squarks of the first
two generations this criterion is almost automatically fulfilled for any
model of interest. Furthermore, the neutralino and chargino mass hierarchy 
and mixing character must be such that the squark decays go through
heavier neutralinos or charginos, and the decays of the latter
into a light Higgs and a lighter neutralino or chargino are open. 
Such a scenario is disfavored if the LSP is gaugino-like. In order to generate a
sufficient number of Higgs bosons in the cascade decays, it is also 
advantageous for (at least one of) the gauginos to be heavier than the 
Higgsinos, so that an intermediate Higgsino decay step can be present. 
In the NMSSM such a situation can be realized quite easily if the LSP is
singlino-like.

While the results presented in this paper 
are based on a rather simple-minded analysis,
involving for instance just a fast detector simulation, we nevertheless 
regard them as very encouraging, motivating a further exploration of
the potential for detecting a light non-SM type Higgs in SUSY cascade
decays. In fact, there exists the exciting possibility that the discovery of
a SUSY signal could go hand in hand with the discovery of one or more
Higgs bosons.

\acknowledgments
We are grateful to B.~Fuks and F.~Staub for valuable assistance with generating the {\tt FeynArts} model file used for parts of this work. We also wish to thank T.~Plehn, M.~Spira, and S.~Brensing for illuminating discussions on the QCD corrections to the squark and gluino production cross sections, and A.~Nikitenko, A.~Raspereza, and M.~Schumacher for giving input on various experimental aspects of our analysis. Finally we thank R.~Benbrik, S.~Heinemeyer, M.~Gomez-Bock, and L.~Zeune for interesting and useful discussions. This work was supported by the Collaborative Research Center SFB676 of the DFG, ``Particles, Strings, and the Early Universe''.

\bibliographystyle{JHEP}
\bibliography{nmssmcasc}

\providecommand{\href}[2]{#2}\begingroup\raggedright\begin{thebibliography}{10}

\bibitem{Nilles:1983ge}
H.~P. Nilles, {\it {Supersymmetry, Supergravity and Particle Physics}},  {\em
  Phys.~Rept.} {\bf 110} (1984) 1--162.

\bibitem{Haber:1984rc}
H.~E. Haber and G.~L. Kane, {\it {The Search for Supersymmetry: Probing Physics
  Beyond the Standard Model}},  {\em Phys.~Rept.} {\bf 117} (1985) 75--263.

\bibitem{Ellwanger:2009dp}
U.~Ellwanger, C.~Hugonie, and A.~M. Teixeira, {\it {The Next-to-Minimal
  Supersymmetric Standard Model}},  {\em Phys. Rept.} {\bf 496} (2010) 1--77,
  [\href{http://xxx.lanl.gov/abs/0910.1785}{{\tt arXiv:0910.1785}}].

\bibitem{Maniatis:2009re}
M.~Maniatis, {\it {The Next-to-Minimal Supersymmetric extension of the Standard
  Model reviewed}},  {\em Int.~J.~Mod.~Phys.~A} {\bf 25} (2010) 3505--3602,
  [\href{http://xxx.lanl.gov/abs/0906.0777}{{\tt arXiv:0906.0777}}].

\bibitem{Schael:2006cr}
{ALEPH, DELPHI, L3, OPAL} Collaboration, S.~Schael {\em et.~al.}, {\it
  {Search for neutral MSSM Higgs bosons at LEP}},  {\em Eur.~Phys.~J.~C} {\bf
  47} (2006) 547--587, [\href{http://xxx.lanl.gov/abs/hep-ex/0602042}{{\tt
  hep-ex/0602042}}].

\bibitem{Nakamura:2010zzi}
{Particle Data Group}, K.~Nakamura {\em et.~al.}, {\it
  {Review of particle physics}},  {\em J.~Phys.~G} {\bf 37} (2010) 075021.

\bibitem{Williams:2007dc}
K.~Williams and G.~Weiglein, {\it {Precise predictions for $h_a \to h_b h_c$
  decays in the complex MSSM}},  {\em Phys.~Lett.~B} {\bf 660} (2008) 217--227,
  [\href{http://xxx.lanl.gov/abs/0710.5320}{{\tt arXiv:0710.5320}}].

\bibitem{Williams:2011bu}
K.~E. Williams, H.~Rzehak, and G.~Weiglein, {\it {Higher order corrections to
  Higgs boson decays in the MSSM with complex parameters}},  {\em
  Eur.~Phys.~J.~C} {\bf 71} (2011) 1669,
  [\href{http://xxx.lanl.gov/abs/1103.1335}{{\tt arXiv:1103.1335}}].

\bibitem{Dermisek:2005ar}
R.~Dermisek and J.~F. Gunion, {\it {Escaping the large fine tuning and little
  hierarchy problems in the next to minimal supersymmetric model and $h \to aa$
  decays}},  {\em Phys.~Rev.~Lett.} {\bf 95} (2005) 041801,
  [\href{http://xxx.lanl.gov/abs/hep-ph/0502105}{{\tt hep-ph/0502105}}].

\bibitem{Dermisek:2005gg}
R.~Dermisek and J.~F. Gunion, {\it {Consistency of LEP event excesses with an
  $h \to aa$ decay scenario and low-fine-tuning NMSSM models}},  {\em
  Phys.~Rev.~D} {\bf 73} (2006) 111701,
  [\href{http://xxx.lanl.gov/abs/hep-ph/0510322}{{\tt hep-ph/0510322}}].

\bibitem{Dermisek:2006wr}
R.~Dermisek and J.~F. Gunion, {\it {The NMSSM Close to the R-symmetry Limit and
  Naturalness in $h \to aa$ Decays for $m_a < 2m_b$)}},  {\em Phys.~Rev.~D}
  {\bf 75} (2007) 075019, [\href{http://xxx.lanl.gov/abs/hep-ph/0611142}{{\tt
  hep-ph/0611142}}].

\bibitem{Dermisek:2007yt}
R.~Dermisek and J.~F. Gunion, {\it {The NMSSM Solution to the Fine-Tuning
  Problem, Precision Electroweak Constraints and the Largest LEP Higgs Event
  Excess}},  {\em Phys.~Rev.~D} {\bf 76} (2007) 095006,
  [\href{http://xxx.lanl.gov/abs/0705.4387}{{\tt arXiv:0705.4387}}].

\bibitem{Dermisek:2008uu}
R.~Dermisek and J.~F. Gunion, {\it {Many Light Higgs Bosons in the NMSSM}},
  {\em Phys. Rev.} {\bf D79} (2009) 055014,
  [\href{http://xxx.lanl.gov/abs/0811.3537}{{\tt arXiv:0811.3537}}].

\bibitem{Dermisek:2010mg}
R.~Dermisek and J.~F. Gunion, {\it {New constraints on a light CP-odd Higgs
  boson and related NMSSM Ideal Higgs Scenarios}},  {\em Phys. Rev. D} {\bf 81}
  (2010) 075003, [\href{http://xxx.lanl.gov/abs/1002.1971}{{\tt
  arXiv:1002.1971}}].

\bibitem{Carena:2000ks}
M.~S. Carena, J.~R. Ellis, A.~Pilaftsis, and C.~Wagner, {\it {CP violating MSSM
  Higgs bosons in the light of LEP-2}},  {\em Phys.~Lett.~B} {\bf 495} (2000)
  155--163, [\href{http://xxx.lanl.gov/abs/hep-ph/0009212}{{\tt
  hep-ph/0009212}}].

\bibitem{Buescher:2005re}
V.~Buescher and K.~Jakobs, {\it {Higgs boson searches at hadron colliders}},
  {\em Int.~J.~Mod.~Phys.~A} {\bf 20} (2005) 2523--2602,
  [\href{http://xxx.lanl.gov/abs/hep-ph/0504099}{{\tt hep-ph/0504099}}].

\bibitem{Schumacher:2004da}
M.~Schumacher, {\it {Investigation of the discovery potential for Higgs bosons
  of the minimal supersymmetric extension of the standard model (MSSM) with
  ATLAS}},  [\href{http://xxx.lanl.gov/abs/hep-ph/0410112}{{\tt
  hep-ph/0410112}}].

\bibitem{Accomando:2006ga}
E.~Accomando, A.~Akeroyd, E.~Akhmetzyanova, J.~Albert, A.~Alves, {\em et.~al.},
  {\it {Workshop on CP Studies and Non-Standard Higgs Physics}},
  [\href{http://xxx.lanl.gov/abs/hep-ph/0608079}{{\tt hep-ph/0608079}}].

\bibitem{Akeroyd:2003jp}
A.~Akeroyd, {\it {Searching for a very light Higgs boson at the Tevatron}},
  {\em Phys.~Rev.~D} {\bf 68} (2003) 077701,
  [\href{http://xxx.lanl.gov/abs/hep-ph/0306045}{{\tt hep-ph/0306045}}].

\bibitem{Ghosh:2004cc}
D.~K. Ghosh, R.~Godbole, and D.~Roy, {\it {Probing the CP-violating light
  neutral Higgs in the charged Higgs decay at the LHC}},  {\em Phys.~Lett.~B}
  {\bf 628} (2005) 131--140,
  [\href{http://xxx.lanl.gov/abs/hep-ph/0412193}{{\tt hep-ph/0412193}}].

\bibitem{Cheung:2007sva}
K.~Cheung, J.~Song, and Q.-S. Yan, {\it {Role of h $\to \eta \eta$ in
  Intermediate-Mass Higgs Boson Searches at the Large Hadron Collider}},  {\em
  Phys.~Rev.~Lett.} {\bf 99} (2007) 031801,
  [\href{http://xxx.lanl.gov/abs/hep-ph/0703149}{{\tt hep-ph/0703149}}].

\bibitem{Carena:2007jk}
M.~Carena, T.~Han, G.-Y. Huang, and C.~E. Wagner, {\it {Higgs Signal for $h \to
  aa$ at Hadron Colliders}},  {\em JHEP} {\bf 0804} (2008) 092,
  [\href{http://xxx.lanl.gov/abs/0712.2466}{{\tt arXiv:0712.2466}}].

\bibitem{Fowler:2009ay}
A.~C. Fowler and G.~Weiglein, {\it {Precise Predictions for Higgs Production in
  Neutralino Decays in the Complex MSSM}},  {\em JHEP} {\bf 01} (2010) 108,
  [\href{http://xxx.lanl.gov/abs/0909.5165}{{\tt arXiv:0909.5165}}].

\bibitem{Draper:2009au}
P.~Draper, T.~Liu, and C.~E. Wagner, {\it {Prospects for Higgs Searches at the
  Tevatron and LHC in the MSSM with Explicit CP-violation}},  {\em
  Phys.~Rev.~D} {\bf 81} (2010) 015014,
  [\href{http://xxx.lanl.gov/abs/0911.0034}{{\tt arXiv:0911.0034}}].

\bibitem{Bandyopadhyay:2010tv}
P.~Bandyopadhyay, {\it {Higgs production in CP-violating supersymmetric cascade
  decays: Probing the `open hole' at the Large Hadron Collider}},
  [\href{http://xxx.lanl.gov/abs/1008.3339}{{\tt arXiv:1008.3339}}].

\bibitem{Bandyopadhyay:2011vc}
P.~Bandyopadhyay and K.~Huitu, {\it {Production of two Higgses at the Large
  Hadron Collider in CP-violating supersymmetry}},
  [\href{http://xxx.lanl.gov/abs/1106.5108}{{\tt arXiv:1106.5108}}].

\bibitem{Barate:2003sz}
{LEP Working Group for Higgs boson searches}, R.~Barate {\em
  et.~al.}, {\it {Search for the standard model Higgs boson at LEP}},  {\em
  Phys.~Lett.~B} {\bf 565} (2003) 61--75,
  [\href{http://xxx.lanl.gov/abs/hep-ex/0306033}{{\tt hep-ex/0306033}}].

\bibitem{Ball:2007zza}
{CMS} Collaboration, G.~Bayatian {\em et.~al.}, {\it {CMS technical design
  report, volume II: Physics performance}},  {\em J.~Phys.~G} {\bf 34} (2007)
  995--1579.

\bibitem{Datta:2003iz}
A.~Datta, A.~Djouadi, M.~Guchait, and F.~Moortgat, {\it {Detection of MSSM
  Higgs bosons from supersymmetric particle cascade decays at the LHC}},  {\em
  Nucl.~Phys.~B} {\bf 681} (2004) 31--64,
  [\href{http://xxx.lanl.gov/abs/hep-ph/0303095}{{\tt hep-ph/0303095}}].

\bibitem{Huitu:2008sa}
K.~Huitu, R.~Kinnunen, J.~Laamanen, S.~Lehti, S.~Roy, {\em et.~al.}, {\it
  {Search for Higgs Bosons in SUSY Cascades in CMS and Dark Matter with
  Non-universal Gaugino Masses}},  {\em Eur.~Phys.~J.~C} {\bf 58} (2008)
  591--608, [\href{http://xxx.lanl.gov/abs/0808.3094}{{\tt arXiv:0808.3094}}].

\bibitem{Gori:2011hj}
S.~Gori, P.~Schwaller, and C.~E. Wagner, {\it {Search for Higgs Bosons in SUSY
  Cascade Decays and Neutralino Dark Matter}},  {\em Phys.~Rev.~D} {\bf 83}
  (2011) 115022, [\href{http://xxx.lanl.gov/abs/1103.4138}{{\tt
  arXiv:1103.4138}}].

\bibitem{Kribs:2009yh}
G.~D. Kribs, A.~Martin, T.~S. Roy, and M.~Spannowsky, {\it {Discovering the
  Higgs Boson in New Physics Events using Jet Substructure}},  {\em
  Phys.~Rev.~D} {\bf 81} (2010) 111501,
  [\href{http://xxx.lanl.gov/abs/0912.4731}{{\tt arXiv:0912.4731}}].

\bibitem{Kribs:2010hp}
G.~D. Kribs, A.~Martin, T.~S. Roy, and M.~Spannowsky, {\it {Discovering Higgs
  Bosons of the MSSM using Jet Substructure}},  {\em Phys.~Rev.~D} {\bf 82}
  (2010) 095012, [\href{http://xxx.lanl.gov/abs/1006.1656}{{\tt
  arXiv:1006.1656}}].

\bibitem{Cheung:2008rh}
K.~Cheung and T.-J. Hou, {\it {Light Pseudoscalar Higgs boson in Neutralino
  Decays in the Next-to-Minimal Supersymmetric Standard Model}},  {\em
  Phys.~Lett.~B} {\bf 674} (2009) 54--58,
  [\href{http://xxx.lanl.gov/abs/0809.1122}{{\tt arXiv:0809.1122}}].

\bibitem{Skands:2003cj}
P.~Z. Skands {\em et.~al.}, {\it {SUSY Les Houches Accord: Interfacing SUSY
  Spectrum Calculators, Decay Packages, and Event Generators}},  {\em JHEP}
  {\bf 07} (2004) 036, [\href{http://xxx.lanl.gov/abs/hep-ph/0311123}{{\tt
  hep-ph/0311123}}].

\bibitem{SLHA2}
B.~Allanach, C.~Balazs, G.~Belanger, M.~Bernhardt, F.~Boudjema, {\em et.~al.},
  {\it {SUSY Les Houches Accord 2}},  {\em Comput.~Phys.~Commun.} {\bf 180}
  (2009) 8--25, [\href{http://xxx.lanl.gov/abs/0801.0045}{{\tt
  arXiv:0801.0045}}].

\bibitem{Djouadi:2008uw}
A.~Djouadi, M.~Drees, U.~Ellwanger, R.~Godbole, C.~Hugonie, {\em et.~al.}, {\it
  {Benchmark scenarios for the NMSSM}},  {\em JHEP} {\bf 0807} (2008) 002,
  [\href{http://xxx.lanl.gov/abs/0801.4321}{{\tt arXiv:0801.4321}}].

\bibitem{Ellwanger:1993hn}
U.~Ellwanger, {\it {Radiative corrections to the neutral Higgs spectrum in
  supersymmetry with a gauge singlet}},  {\em Phys.~Lett.~B} {\bf 303} (1993)
  271--276, [\href{http://xxx.lanl.gov/abs/hep-ph/9302224}{{\tt
  hep-ph/9302224}}].

\bibitem{Elliott:1993ex}
T.~Elliott, S.~King, and P.~White, {\it {Supersymmetric Higgs bosons at the
  limit}},  {\em Phys.~Lett.~B} {\bf 305} (1993) 71--77,
  [\href{http://xxx.lanl.gov/abs/hep-ph/9302202}{{\tt hep-ph/9302202}}].

\bibitem{Elliott:1993uc}
T.~Elliott, S.~King, and P.~White, {\it {Squark contributions to Higgs boson
  masses in the next-to-minimal supersymmetric standard model}},  {\em
  Phys.~Lett.~B} {\bf 314} (1993) 56--63,
  [\href{http://xxx.lanl.gov/abs/hep-ph/9305282}{{\tt hep-ph/9305282}}].

\bibitem{Elliott:1993bs}
T.~Elliott, S.~King, and P.~White, {\it {Radiative corrections to Higgs boson
  masses in the next-to-minimal supersymmetric Standard Model}},  {\em
  Phys.~Rev.~D} {\bf 49} (1994) 2435--2456,
  [\href{http://xxx.lanl.gov/abs/hep-ph/9308309}{{\tt hep-ph/9308309}}].

\bibitem{Pandita:1993hx}
P.~Pandita, {\it {One loop radiative corrections to the lightest Higgs scalar
  mass in nonminimal supersymmetric Standard Model}},  {\em Phys.~Lett.~B} {\bf
  318} (1993) 338--346.

\bibitem{Ellwanger:2005fh}
U.~Ellwanger and C.~Hugonie, {\it {Yukawa induced radiative corrections to the
  lightest Higgs boson mass in the NMSSM}},  {\em Phys.~Lett.~B} {\bf 623}
  (2005) 93--103, [\href{http://xxx.lanl.gov/abs/hep-ph/0504269}{{\tt
  hep-ph/0504269}}].

\bibitem{Degrassi:2009yq}
G.~Degrassi and P.~Slavich, {\it {On the radiative corrections to the neutral
  Higgs boson masses in the NMSSM}},  {\em Nucl.~Phys.~B} {\bf 825} (2010)
  119--150, [\href{http://xxx.lanl.gov/abs/0907.4682}{{\tt arXiv:0907.4682}}].

\bibitem{Ellwanger:2004xm}
U.~Ellwanger, J.~F. Gunion, and C.~Hugonie, {\it {NMHDECAY: A Fortran code for
  the Higgs masses, couplings and decay widths in the NMSSM}},  {\em JHEP} {\bf
  02} (2005) 066, [\href{http://xxx.lanl.gov/abs/hep-ph/0406215}{{\tt
  hep-ph/0406215}}].

\bibitem{Ellwanger:2005dv}
U.~Ellwanger and C.~Hugonie, {\it {NMHDECAY 2.0: An Updated program for
  sparticle masses, Higgs masses, couplings and decay widths in the NMSSM}},
  {\em Comput.~Phys.~Commun.} {\bf 175} (2006) 290--303,
  [\href{http://xxx.lanl.gov/abs/hep-ph/0508022}{{\tt hep-ph/0508022}}].

\bibitem{Belanger:2005kh}
G.~Belanger, F.~Boudjema, C.~Hugonie, A.~Pukhov, and A.~Semenov, {\it {Relic
  density of dark matter in the NMSSM}},  {\em JCAP} {\bf 0509} (2005) 001,
  [\href{http://xxx.lanl.gov/abs/hep-ph/0505142}{{\tt hep-ph/0505142}}].

\bibitem{Mahmoudi:2010xp}
F.~Mahmoudi, J.~Rathsman, O.~St{\aa}l, and L.~Zeune, {\it {Light Higgs bosons
  in phenomenological NMSSM}},  {\em Eur.~Phys.~J.~C} {\bf 71} (2011) 1608,
  [\href{http://xxx.lanl.gov/abs/1012.4490}{{\tt arXiv:1012.4490}}].

\bibitem{Eriksson:2008cx}
D.~Eriksson, F.~Mahmoudi, and O.~St{\aa}l, {\it {Charged Higgs bosons in
  Minimal Supersymmetry: Updated constraints and experimental prospects}},
  {\em JHEP} {\bf 11} (2008) 035,
  [\href{http://xxx.lanl.gov/abs/0808.3551}{{\tt arXiv:0808.3551}}].

\bibitem{Stockinger:2006zn}
D.~St{\"o}ckinger, {\it {The Muon Magnetic Moment and Supersymmetry}},  {\em
  J.~Phys.~G} {\bf 34} (2007) R45--R92,
  [\href{http://xxx.lanl.gov/abs/hep-ph/0609168}{{\tt hep-ph/0609168}}].

\bibitem{Ellwanger:2010nf}
U.~Ellwanger, {\it {Enhanced di-photon Higgs signal in the Next-to-Minimal
  Supersymmetric Standard Model}},  {\em Phys.~Lett.~B} {\bf 698} (2011)
  293--296, [\href{http://xxx.lanl.gov/abs/1012.1201}{{\tt arXiv:1012.1201}}].

\bibitem{Ellwanger:2001iw}
U.~Ellwanger, J.~F. Gunion, and C.~Hugonie, {\it {Establishing a no-lose
  theorem for NMSSM Higgs boson discovery at the LHC}},
  [\href{http://xxx.lanl.gov/abs/hep-ph/0111179}{{\tt hep-ph/0111179}}].

\bibitem{Ellwanger:2003jt}
U.~Ellwanger, J.~F. Gunion, C.~Hugonie, and S.~Moretti, {\it {Towards a no-lose
  theorem for NMSSM Higgs discovery at the LHC}},
  [\href{http://xxx.lanl.gov/abs/hep-ph/0305109}{{\tt hep-ph/0305109}}].

\bibitem{Ellwanger:2004gz}
U.~Ellwanger, J.~F. Gunion, C.~Hugonie, and S.~Moretti, {\it {NMSSM Higgs
  discovery at the LHC}},  [\href{http://xxx.lanl.gov/abs/hep-ph/0401228}{{\tt
  hep-ph/0401228}}].

\bibitem{Forshaw:2007ra}
J.~R. Forshaw, J.~F. Gunion, L.~Hodgkinson, A.~Papaefstathiou, and A.~D.
  Pilkington, {\it {Reinstating the 'no-lose' theorem for NMSSM Higgs discovery
  at the LHC}},  {\em JHEP} {\bf 04} (2008) 090,
  [\href{http://xxx.lanl.gov/abs/0712.3510}{{\tt arXiv:0712.3510}}].

\bibitem{Belyaev:2008gj}
A.~Belyaev, S.~Hesselbach, S.~Lehti, S.~Moretti, A.~Nikitenko, {\em et.~al.},
  {\it {The Scope of the 4 tau Channel in Higgs-strahlung and Vector Boson
  Fusion for the NMSSM No-Lose Theorem at the LHC}},
  [\href{http://xxx.lanl.gov/abs/0805.3505}{{\tt arXiv:0805.3505}}].

\bibitem{Choi:2004zx}
S.~Choi, D.~J.~Miller, and P.~Zerwas, {\it {The Neutralino sector of the
  next-to-minimal supersymmetric standard model}},  {\em Nucl.~Phys.~B} {\bf
  711} (2005) 83--111, [\href{http://xxx.lanl.gov/abs/hep-ph/0407209}{{\tt
  hep-ph/0407209}}].

\bibitem{Liebler:2010bi}
S.~Liebler and W.~Porod, {\it {Electroweak corrections to Neutralino and
  Chargino decays into a W-boson in the (N)MSSM}},  {\em Nucl.~Phys.~B} {\bf
  849} (2011) 213--249, [\href{http://xxx.lanl.gov/abs/1011.6163}{{\tt
  arXiv:1011.6163}}].

\bibitem{Feynarts}
T.~Hahn, {\it {Generating Feynman diagrams and amplitudes with FeynArts 3}},
  {\em Comput.~Phys.~Commun.} {\bf 140} (2001) 418--431,
  [\href{http://xxx.lanl.gov/abs/hep-ph/0012260}{{\tt hep-ph/0012260}}].

\bibitem{FormCalc}
T.~Hahn and M.~Perez-Victoria, {\it {Automatized one-loop calculations in four
  and D dimensions}},  {\em Comput.~Phys.~Commun.} {\bf 118} (1999) 153--165,
  [\href{http://xxx.lanl.gov/abs/hep-ph/9807565}{{\tt hep-ph/9807565}}].

\bibitem{Christensen:2008py}
N.~D. Christensen and C.~Duhr, {\it {FeynRules - Feynman rules made easy}},
  {\em Comput.~Phys.~Commun.} {\bf 180} (2009) 1614--1641,
  [\href{http://xxx.lanl.gov/abs/0806.4194}{{\tt arXiv:0806.4194}}].

\bibitem{Staub:2009bi}
F.~Staub, {\it {From Superpotential to Model Files for FeynArts and
  CalcHep/CompHep}},  {\em Comput.~Phys.~Commun.} {\bf 181} (2010) 1077--1086,
  [\href{http://xxx.lanl.gov/abs/0909.2863}{{\tt arXiv:0909.2863}}].

\bibitem{Beenakker:1996ch}
W.~Beenakker, R.~Hopker, M.~Spira, and P.~M. Zerwas, {\it {Squark and gluino
  production at hadron colliders}},  {\em Nucl.~Phys.~B} {\bf 492} (1997)
  51--103, [\href{http://xxx.lanl.gov/abs/hep-ph/9610490}{{\tt
  hep-ph/9610490}}].

\bibitem{Pumplin:2002vw}
J.~Pumplin {\em et.~al.}, {\it {New generation of parton distributions with
  uncertainties from global QCD analysis}},  {\em JHEP} {\bf 07} (2002) 012,
  [\href{http://xxx.lanl.gov/abs/hep-ph/0201195}{{\tt hep-ph/0201195}}].

\bibitem{Beenakker:1997ut}
W.~Beenakker, M.~Kramer, T.~Plehn, M.~Spira, and P.~M. Zerwas, {\it {Stop
  production at hadron colliders}},  {\em Nucl.~Phys.~B} {\bf 515} (1998)
  3--14, [\href{http://xxx.lanl.gov/abs/hep-ph/9710451}{{\tt hep-ph/9710451}}].

\bibitem{Aad:2011hh}
{ATLAS} Collaboration, G.~Aad {\em et.~al.}, {\it {Search for supersymmetry
  using final states with one lepton, jets, and missing transverse momentum
  with the ATLAS detector in $\sqrt{s} = 7$~TeV pp}},  {\em Phys.~Rev.~Lett.}
  {\bf 106} (2011) 131802, [\href{http://xxx.lanl.gov/abs/1102.2357}{{\tt
  arXiv:1102.2357}}].

\bibitem{daCosta:2011qk}
{ATLAS} Collaboration, J.~B.~G. da~Costa {\em et.~al.}, {\it {Search for
  squarks and gluinos using final states with jets and missing transverse
  momentum with the ATLAS detector in $\sqrt{s}$ = 7 TeV proton-proton
  collisions}},  {\em Phys.~Lett.~B} {\bf 701} (2011) 186--203,
  [\href{http://xxx.lanl.gov/abs/1102.5290}{{\tt arXiv:1102.5290}}].

\bibitem{Aad:2011ks}
{ATLAS} Collaboration, G.~Aad {\em et.~al.}, {\it {Search for supersymmetry
  in pp collisions at $\sqrt{s} = 7$~TeV in final states with missing
  transverse momentum and b-jets}},
  [\href{http://xxx.lanl.gov/abs/1103.4344}{{\tt arXiv:1103.4344}}].

\bibitem{Collaboration:2011xk}
{ATLAS} Collaboration, {\it {Search for an excess of events with an
  identical flavour lepton pair and significant missing transverse momentum in
  $\sqrt{s} = 7$~TeV proton-proton collisions with the ATLAS detector}},
  [\href{http://xxx.lanl.gov/abs/1103.6208}{{\tt arXiv:1103.6208}}].

\bibitem{Aad:2011xm}
{ATLAS} Collaboration, G.~Aad {\em et.~al.}, {\it {Search for
  supersymmetric particles in events with lepton pairs and large missing
  transverse momentum in $\sqrt{s}= 7$ TeV proton-proton collisions with the
  ATLAS experiment}},  [\href{http://xxx.lanl.gov/abs/1103.6214}{{\tt
  arXiv:1103.6214}}].

\bibitem{Khachatryan:2011tk}
{CMS} Collaboration, V.~Khachatryan {\em et.~al.}, {\it {Search for
  Supersymmetry in pp Collisions at 7 TeV in Events with Jets and Missing
  Transverse Energy}},  {\em Phys.~Lett.~B} {\bf 698} (2011) 196--218,
  [\href{http://xxx.lanl.gov/abs/1101.1628}{{\tt arXiv:1101.1628}}].

\bibitem{Chatrchyan:2011nx}
{CMS} Collaboration, S.~Chatrchyan {\em et.~al.}, {\it {Search for Neutral
  MSSM Higgs Bosons Decaying to Tau Pairs in pp Collisions at $\sqrt{s}$=7
  TeV}}, [\href{http://xxx.lanl.gov/abs/1104.1619}{{\tt arXiv:1104.1619}}].

\bibitem{Chatrchyan:2011ah}
{CMS} Collaboration, S.~Chatrchyan {\em et.~al.}, {\it {Search for
  supersymmetry in events with a lepton, a photon, and large missing transverse
  energy in pp collisions at $\sqrt{s} = 7$~TeV}},
  [\href{http://xxx.lanl.gov/abs/1105.3152}{{\tt arXiv:1105.3152}}].

\bibitem{Chatrchyan:2011bj}
{CMS} Collaboration, S.~Chatrchyan {\em et.~al.}, {\it {Search for
  Supersymmetry in Events with b Jets and Missing Transverse Momentum at the
  LHC}}, [\href{http://xxx.lanl.gov/abs/1106.3272}{{\tt arXiv:1106.3272}}].

\bibitem{ATLAS-CONF-2011-086}
{ATLAS} Collaboration, {\it {Search for squarks and gluinos using final
  states with jets and missing transverse momentum with the ATLAS detector in
  $\sqrt{s}=7$~TeV proton-proton collisions}},
  \href{http://cdsweb.cern.ch/record/1356194?ln=en}{{\tt
  ATLAS-CONF-2011-086}}.

\bibitem{CMS-PAS-SUS-11-003}
{CMS} Collaboration, {\it Search for supersymmetry in all-hadronic events
  with $\alpha_t$},  \href{http://cdsweb.cern.ch/record/1370596?ln=en}{{\tt
  CMS-PAS-SUS-11-003}}.

\bibitem{Alwall:2007st}
J.~Alwall {\em et.~al.}, {\it {MadGraph/MadEvent v4: The New Web Generation}},
  {\em JHEP} {\bf 09} (2007) 028,
  [\href{http://xxx.lanl.gov/abs/0706.2334}{{\tt arXiv:0706.2334}}].

\bibitem{Sjostrand:2006za}
T.~Sj{\"o}strand, S.~Mrenna, and P.~Z. Skands, {\it {PYTHIA 6.4 Physics and
  Manual}},  {\em JHEP} {\bf 05} (2006) 026,
  [\href{http://xxx.lanl.gov/abs/hep-ph/0603175}{{\tt hep-ph/0603175}}].

\bibitem{Ovyn:2009tx}
S.~Ovyn, X.~Rouby, and V.~Lemaitre, {\it {Delphes, a framework for fast
  simulation of a generic collider experiment}},
  [\href{http://xxx.lanl.gov/abs/0903.2225}{{\tt arXiv:0903.2225}}].

\bibitem{Cacciari:2008gp}
M.~Cacciari, G.~P. Salam, and G.~Soyez, {\it {The anti-$k_t$ jet clustering
  algorithm}},  {\em JHEP} {\bf 04} (2008) 063,
  [\href{http://xxx.lanl.gov/abs/0802.1189}{{\tt arXiv:0802.1189}}].

\bibitem{Aad:2009wy}
{ATLAS} Collaboration, G.~Aad {\em et.~al.}, {\it {Expected Performance of
  the ATLAS Experiment - Detector, Trigger and Physics}},
  [\href{http://xxx.lanl.gov/abs/0901.0512}{{\tt arXiv:0901.0512}}].

\bibitem{Aliev:2010zk}
M.~Aliev {\em et.~al.}, {\it {-- HATHOR -- HAdronic Top and Heavy quarks crOss
  section calculatoR}},  {\em Comput.~Phys.~Commun.} {\bf 182} (2011)
  1034--1046, [\href{http://xxx.lanl.gov/abs/1007.1327}{{\tt
  arXiv:1007.1327}}].

\bibitem{Martin:2009iq}
A.~D. Martin, W.~J. Stirling, R.~S. Thorne, and G.~Watt, {\it {Parton
  distributions for the LHC}},  {\em Eur.~Phys.~J.~C} {\bf 63} (2009) 189--285,
  [\href{http://xxx.lanl.gov/abs/0901.0002}{{\tt arXiv:0901.0002}}].

\end{thebibliography}\endgroup

\end{document}